\documentclass[english,floatfix,prd,superscriptaddress,twocolumn,eqsecnum,nofootinbib,longbibliography]{revtex4-2}
\usepackage{babel}
\usepackage{amsmath}
\usepackage{amssymb,bbold}
\usepackage{graphicx}
\usepackage{amsthm}
 \usepackage[hyperfootnotes=false]{hyperref}
 \usepackage[normalem]{ulem}

\usepackage{color,xcolor,verbatim}
\usepackage{graphicx}% Include figure files
\usepackage[caption=false]{subfig}
\usepackage{soul}

\newcommand{\lamslm}{{}_s\lambda_{\ell m}}
\newcommand{\del}[1][z]{\partial_{#1}}
\newcommand{\indmode}{\ell m\omega} 
\newcommand{\ik}{\mathrm{k}}

\newcommand{\oPSn}[1]{ \omega_{\text{PS}}} 
\newcommand{\oNEn}[1]{ \omega_{\text{NE}}}
\newcommand{\odSn}[1]{ \omega_{\text{dS}}}

\newcommand{\bo}[1]{\bar\omega_{#1}}
\newcommand{\boN}[1]{\omega_{#1}}
\newcommand{\oq}[1]{\omega_{#1,q}}

\newcommand{\indmm}{\mathrm{n}}
\newcommand{\rmm}{r_{\indmm}}

\newcommand{\phio}{\bar\varphi}

\newcommand{\Dr}{\Delta_{r}}
\newcommand{\OdS}{\Omega_{c,0}} 
\newcommand{\kdS}{\kappa_{c,0}}

\newcommand{\rc}{r_c}
\newcommand{\phic}{\phi_c}

\newcommand{\lambdaq}{\lambda_q}

\newcommand{\ph}{\text{ph}}
\newcommand{\LyapLambda}{\varLambda_\ph}

\newcommand{\qcrit}{\bar{q}}
\newcommand{\acrit}{\bar{a}}
\newcommand{\acritq}{\bar{a}}

\newcommand{\R}{R_s} 
\newcommand{\Ang}{S_s}

\newcommand{\an}[1]{\alpha_ {#1}}
\newcommand{\bn}[1]{\beta_ {#1}}
\newcommand{\cn}[1]{\gamma_ {#1}}

\newcommand{\PRDvtwo}[1]{#1}

\newcommand{\oQNM}{\omega_{\ell mn}}

\begin{document}

\title{Glimpses of Violation of Strong Cosmic Censorship  in Rotating Black Holes}

\author{Marc Casals}
\email{marc.casals@uni-leipzig.de, mcasals@cbpf.br, marc.casals@ucd.ie}
\affiliation{Institut f\"ur Theoretische Physik, Universit\"at Leipzig,  Br\"uderstra\ss e  16, 04103 Leipzig, Germany.}
\affiliation{Centro Brasileiro de Pesquisas F\'isicas (CBPF), Rio de Janeiro, 
CEP 22290-180, 
Brazil.}
\affiliation{School of Mathematics and Statistics, University College Dublin, Belfield, Dublin 4, Ireland.}
\author{C\'assio I.~S.~Marinho}%
\email{marinho@cbpf.br}
\affiliation{Centro Brasileiro de Pesquisas F\'isicas (CBPF), Rio de Janeiro, 
CEP 22290-180, 
Brazil.}

\begin{abstract}
Rotating and/or charged black hole spacetimes possess a Cauchy horizon, beyond which  Einstein's equations of General Relativity cease to be deterministic.
This led to the formulation of the Strong Cosmic Censorship conjecture that such horizons become irregular under field perturbations.
We consider linear field perturbations of rotating and electrically-charged  (Kerr-Newman-de Sitter) black holes in a \PRDvtwo{u}niverse with a positive cosmological constant.
By calculating the quasinormal modes for scalar and fermion fields, we provide evidence for the existence of weak solutions to Einstein's equations across the Cauchy horizon in the nearly-extremal regime.
We thus provide,  for the first time, evidence for violation of Strong Cosmic Censorship in  ($4$-dimensional) rotating black hole spacetimes.
\end{abstract}

\maketitle

%---------------------------------------------------------------------------------------------------------
%---------------------------------------------------------------------------------------------------------

\section{Introduction}

%Black hole spacetimes are exact solutions of Einstein's equations of General Relativity.
\PRDvtwo{We consider black holes that are exact solutions of Einstein's equations of General Relativity.} When a black hole (in isolation) is rotating and/or is electrically charged, there is a null hypersurface in its inside, called the Cauchy horizon, beyond which the Cauchy value problem is not well-posed: Einstein's equations cease to be deterministic.
However, black holes in Nature typically form from the gravitational collapse of matter and they are not in isolation. 
It is therefore important to ascertain whether the Cauchy horizon continues to exist as a regular hypersurface when considering matter field perturbations of black hole spacetimes.

\PRDvtwo{Penrose conjectured \cite{1979grec.conf..581P} that Cauchy horizons become irregular when perturbed, so that Einstein’s equations are deterministic inside, as well as outside, black holes that exist in Nature. Such conjecture is known as strong cosmic censorship (SCC). More
specifically, Christodoulou’s formulation of  SCC \cite{Christodoulou:2008nj} consists in the conjecture that Einstein’s equations do not admit weak solutions -- metric solutions with locally square integrable derivatives -- across a Cauchy horizon formed from generic initial data.
In this paper we shall consider linear perturbations of a background  spacetime by matter fields which, in their turn, source higher order metric perturbations. In that case, SCC requires that the stress-energy tensors of the matter fields are not locally integrable. We note that one may alternatively view the matter field perturbations as a proxy for the gravitational perturbations.}
%\sout{Given the notorious difficulty in solving Einstein's nonlinear and coupled equations, it is useful and common to, instead, analyze the equations of {\it linear}  field perturbations which, in their turn, source higher order metric perturbations. The linear perturbations can be gravitational or some matter field's\PRDvtwo{, }either as an analogue of the linear gravitational perturbations or of physical interest in their own right.}

For black holes in a \PRDvtwo{u}niverse with zero \PRDvtwo{c}osmological constant ($\Lambda=0$), strong evidence has been provided that SCC is respected~\cite{poisson1989inner,Ori:1991zz,dafermos2005interior,luk2017strong,luk2017proof,ori1992structure,DafermosLuk2017,Dafermos:2015bzz}.
This is essentially due to the fact that field  perturbations, even though they decay outside the black hole, they do not do so fast enough to compensate for the blueshift that they experience as they approach the Cauchy horizon, so that the field energy blows up at the Cauchy horizon -- a mechanism called {\it mass inflation}~\cite{poisson1990internal}. 

We shall henceforth consider instead black holes in a \PRDvtwo{u}niverse with {\it positive} cosmological constant ($\Lambda>0$), i.e., in a de Sitter (dS) \PRDvtwo{u}niverse. In this case,  field perturbations may become ``dispersed" enough so that when they reach the Cauchy horizon they might not be \PRDvtwo{``}strong\PRDvtwo{''} enough to make it irregular. 

A way of measuring the \PRDvtwo{strength} of a linear field perturbation is via the so-called quasinormal modes (QNMs; see, e.g.,~\cite{Kokkotas:Schmidt:1999,Berti:2009kk} for reviews).
These are field modes with a complex frequency $\oQNM$, whose (negative) imaginary part determines the decay rate of the mode outside the black hole.
Indeed, a key quantity is $\beta$\PRDvtwo{, defined as the infimum of $-\text{Im}(\oQNM)/\kappa_-$ over all QNMs}, where $\kappa_-$ is the surface gravity of the Cauchy horizon. It has been shown \PRDvtwo{in Refs.~\cite{Hintz:2015jkj,PhysRevD.97.104060,Dias2018,Destounis_2019,PhysRevD.98.104007,Ge_2019,Liu:2019rbq,mostafizur2020validity} and in Sec.~\ref{sec:beta} below, in various black hole settings, that %\sout{various physical black hole settings},
%\sout{(which do not include our setting, but in Sec.~\ref{sec:beta} we extend the result to our setting)} 
$\beta>1/2$ corresponds, for linear scalar, fermion and electromagnetic field perturbations, to their  stress-energy tensors being locally integrable at the Cauchy horizon and, for linear gravitational perturbations, to their derivatives being locally square integrable at the Cauchy horizon. Thus,  $\beta>1/2$ corresponds to violation of the linear version of SCC for either matter or gravitational fields}.
Furthermore, $\beta\PRDvtwo{\geq}1$ for spin-0 fields and $\beta\PRDvtwo{\geq}3/2$ for spin-1/2 correspond, via backreaction \PRDvtwo{through the Einstein equations}, to boundedness of the curvature invariants on the Cauchy horizon. 
\PRDvtwo{We note that} nonlinear results (e.g.,~\cite{hintz2018,Hintz2016NonlinearSO,Costa2018}) suggest that conclusions about SCC drawn at the linear level carry over to the nonlinear level\footnote{Violation of SCC in the {\it non}linear setup seems to be inconclusive with the current state-of-the-art techniques even within spherical symmetry (see, e.g., Ref~\cite{PhysRevD.103.104043,zhang2019strong,luna2019strong}).}.

In the case of charged non-rotating black holes in dS (Reissner-Nordstr\"om-dS, RNdS), it has been  shown%~\sout{[16-19,28-32] (see also~\cite{HOD2019636})}
~\cite{PhysRevD.41.403,Cardoso:2017soq,PhysRevD.98.104007,PhysRevD.98.124025,Dias:2018ufh,Guo:2019tjy,Ge_2019,Destounis_2019,Dias2018} that $\beta>1/2$ is possible for nearly-maximal charge (i.e., near extremality). 
\PRDvtwo{Thus, there is violation of the linear version of SCC in RNdS.}

Black holes in Nature, in their turn, are expected (e.g.~\cite{gammie2004black}) to be  {\it rotating} and so it is particularly important to investigate whether the Cauchy horizons of rotating black holes become irregular under perturbations.
So far, all investigations for rotating black holes were in the linear scenario and they all found that SCC is preserved ($\beta<1/2$). That was shown for neutral rotating black holes in dS (Kerr-dS)
under scalar and gravitational field perturbations in~\cite{PhysRevD.97.104060} and in our App.~\ref{sec:KdS} for  neutrino field perturbations \PRDvtwo{(which invalidates the conclusion in Ref.~\cite{mostafizur2020validity} that there is violation of SCC by neutrino fields in Kerr-dS)}. 
When  including black hole charge as well as rotation (Kerr-Newman-dS, KNdS), preservation of SCC was shown in~\cite{HOD2018221}, where only angular momentum above a certain value was considered, and in~\cite{Rahman:2018oso}, where only a limited range of parameters was investigated.
In this work, on the other hand, we provide strong evidence that violation of SCC can actually be achieved in KNdS black holes, \PRDvtwo{for nearly-maximal charge (similarly to their static counterpart) and in regions of parameter space outside those considered in~\cite{HOD2018221,Rahman:2018oso}}.
Specifically, we calculate the QNMs of massless, both neutral and charged, scalar and fermion  field perturbations of KNdS. We then show that $\beta>1/2$ is possible. 
In fact, for zero or small field charge, we show that even $\beta>1$ for scalar fields and $\beta>3/2$ for fermion fields are possible.

We also report that, in numerical searches, we found no unstable modes (i.e., modes with $\text{Im}\left(\oQNM\right)>0$, exponentially growing in time)  for the KNdS spacetime parameters that we considered in this paper.

The rest of this paper is organized as follows. In Sec.\ref{sec:perturbations}
we introduce KNdS spacetimes,  spin-field perturbations of KNdS and  QNMs.
We describe the various families of QNMs in KNdS in Sec.~\ref{sec:families}, providing analytic expressions that approximate the frequencies of the various families.
In Sec.~\ref{sec:num meth} we describe our numerical method. We then present our results for $\beta$ across various regions of parameter space and the consequences for SCC  in Sec.~\ref{sec:SCC}.
We conclude the main body of the paper with a discussion in Sec.~\ref{sec:Discussion}.
Finally, we have three appendixes: in App.~\ref{sec:KdS} we show that SCC is preserved in Kerr-dS; 
in App.~\ref{sec:geo_approx} we study circular photon orbits in KNdS, which are the basis for the analytic expression for the so-called photon-sphere family of modes;  finally, in App.~\ref{sec:app:Dirac}, we introduce the main quantities needed for obtaining the conditions for $\beta$ for massless charged fermion fields perturbations.

We choose units such that $G=c=\hbar=1$ and metric signature $(-,+,+,+)$. 

%----------------------------------------------------
%----------------------------------------------------

\section{Perturbations of \texorpdfstring{KN\lowercase{d}S}{KNDS} black holes}\label{sec:perturbations}
\subsection{KNdS spacetime}

The line-element of KNdS  spacetimes  can be written as~\cite{carter1968hamilton,Gibbons:Hawking:1977}
%\begin{widetext}
%\begin{equation}\label{eq:KNdSmetric}
%ds^2 = -\frac{\Dr}{\rho^2\Xi^2}\left(dt-a\sin^{2}\theta %d\varphi\right)^2 +\frac{\Delta_{\theta}\sin^{2}\theta}{\rho^2\Xi^2}\left(a\, dt-(r^2+a^{2})d\varphi\right)^2 +\rho^2 \left(\frac{ d\theta^2}{\Delta_{\theta}} +\frac{dr^2}{\Dr}\right),
%\end{equation}
%\end{widetext}

\begin{align}\label{eq:KNdSmetric}
& ds^2 = -\frac{\Dr}{\rho^2\Xi^2}\left(dt-a\sin^{2}\theta d\varphi\right)^2 +    \nonumber\\
& \frac{\Delta_{\theta}\sin^{2}\theta}{\rho^2\Xi^2}\left(a\, dt-(r^2+a^{2})d\varphi\right)^2 +\rho^2 \left(\frac{ d\theta^2}{\Delta_{\theta}} +
    \frac{dr^2}{\Dr}\right),
\end{align}
where
$t,r\in \mathbb{R}$, $\theta \in [0,\pi]$,
$\varphi\in [0,2\pi)$,
\begin{align}
  \label{eq:170}
  \quad \rho^{2} &\equiv r^{2} + a^{2}\cos^{2}\theta, &\quad \Delta_{\theta} &\equiv 1 + \alpha\cos^{2}\theta,
\end{align}
\begin{equation}
    \Dr(r) \equiv (r^{2}+a^{2})\left(1-\frac{r^{2}}{L^{2}}\right)-2Mr+Q^2,
\end{equation}
with $\alpha\equiv \Xi -1\equiv a^2/L^2$, $L \equiv \sqrt{3/\Lambda}$ and cosmological constant $\Lambda>0$.
In the case that the spacetime is that of a black hole, $M$ is the mass,  $a$ is the angular momentum per unit mass and  $Q$ is the charge.
The electromagnetic potential associated with the charge $Q$ is given by
\begin{equation}\label{eq:eletromag_pot}
\boldsymbol{A}=-\frac{Qr}{\Xi \rho^2}(dt-a\sin^2\theta d\varphi).
\end{equation}
The values of the spacetime parameters $\{a,Q,M,\Lambda\}$ that yield all real roots of $\Dr$, as a polynomial in $r$, correspond to a black hole  spacetime; this root-reality condition is equivalent to the discriminant of $\Dr$ being non-negative. In this black hole case, $\Dr$ \PRDvtwo{admits up to four roots}: an event horizon at the radius $r_+$, a Cauchy horizon at $r_-$, a \PRDvtwo{c}osmological horizon at $\rc$, and an inner \PRDvtwo{c}osmological horizon at $\rmm\equiv-(\rc+r_++r_-)$, where $\rmm< 0 \leq r_- \leq r_+ \leq \rc$. 
Each horizon $r_j$, $j\in \{\indmm,+,-,c\}$, has an associated angular velocity
$\Omega_j$, surface gravity $\kappa_j$ and electric potential $\phi_{j}$ given by:
\begin{align}
\Omega_j&\equiv\frac{a}{(r_{j}^2+a^2)},\quad
\kappa_j\equiv\frac{\left|\Dr'(r_{j})\right|}{2\, \Xi\, (r_{j}^2+a^2)},\nonumber\\
\phi_{j}&\equiv \phi(r_j),
\quad \phi(r)\equiv \frac{Q r}{\Xi \left(r^2+a^2\right) },\nonumber
\end{align}
 where we note that  $\kappa_+\leq \kappa_-$.
Extremal black holes correspond to $r_-=r_+$, which is achieved at some upper bounds $Q=Q_\text{max}$ and $a=a_\text{max}$ for charge and rotation, respectively.
Near the extremal black hole limit, it is $\kappa_+\sim \kappa_-\to 0^+$. There are also the horizon-confluence cases $r_+=\rc$, called rotating Nariai, and  $r_+=r_-=r_c$, called ``ultra-extreme". 

\begin{figure}[!htb]
    \centering
    \includegraphics[width=0.478\textwidth]{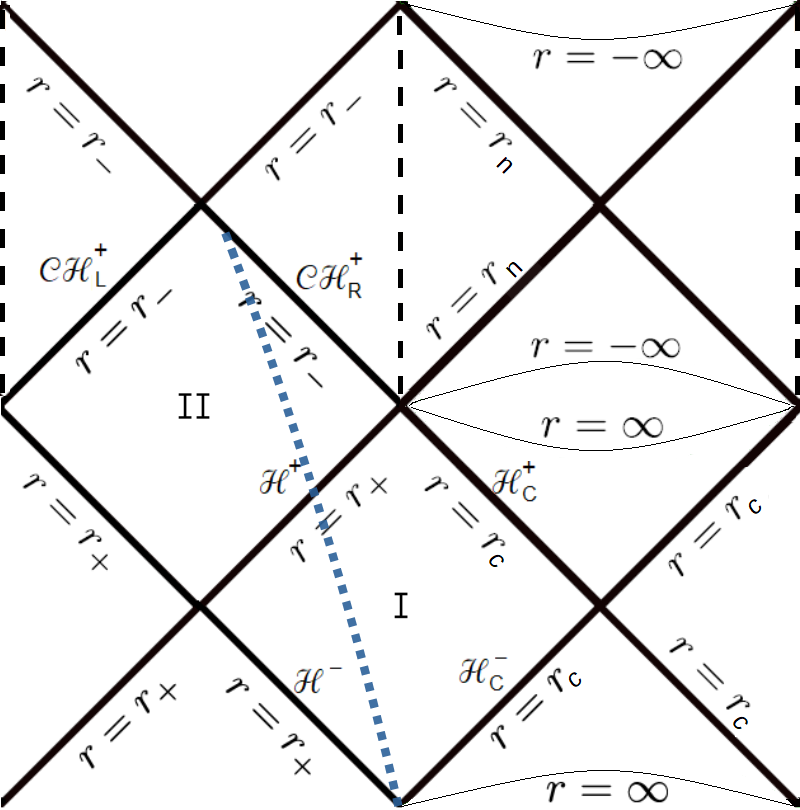}
    \caption{\PRDvtwo{Part of a Carter-Penrose diagram of KNdS spacetime showing the four horizons at $r=r_{\text{n}}$, $r_-$, $r_+$, $r_c$ (solid lines at $45^\circ$), the curvature singularities at $\rho^2=0$ (dashed vertical lines; in the case of $\theta=\pi/2$, the spacetime ends there), the spacelike hypersurfaces at $r=\pm \infty$ (curved thin lines) and a timelike worldline (dotted blue line), which schematically represents the surface of gravitationally-collapsing matter that forms the black hole.}}
    \label{fig:CP-KNdS}
\end{figure}

\PRDvtwo{
These horizons are illustrated in the Carter-Penrose diagram of KNdS in Fig.~\ref{fig:CP-KNdS}.
We note that the (future) Cauchy horizon is composed of two pieces: a right piece $\mathcal{CH}_\text{R}^+$ and a left piece  $\mathcal{CH}_\text{L}^+$.
Furthermore, in the region inside the black hole which is beyond  $\mathcal{CH}_\text{R}^+\cup \mathcal{CH}_\text{L}^+$,
there is: (i) a curvature (ring) singularity at $\rho^2=0$ (and so at $r=0$ and $\theta=\pi/2$; see the dashed vertical lines in Fig.~\ref{fig:CP-KNdS}); (ii) closed timelike curves.  
In this region, the Cauchy value problem is not well-posed and Einstein equations cease to be deterministic.
In the case that the black hole forms from the gravitational collapse of matter (see the dotted blue line in Fig.~\ref{fig:CP-KNdS}),  (possibly, only part of) the right piece $\mathcal{CH}_\text{R}^+$ is the relevant part of the Cauchy horizon.}

\subsection{Field perturbation equations}

We consider linear field perturbations $\Psi_{|s|}$ of sub-extremal KNdS black holes, where $s$ denotes the spin\footnote{Strictly speaking, $s$ here denotes the helicity of the field but, as a common abuse of language, we shall refer to it as the spin.}
of the field. Spin $s=0$ corresponds to a massless, conformally-coupled scalar field $\Psi_0$ of charge $q$, which satisfies the wave equation \cite{Konoplya:2007zx}:
\begin{equation}\label{eq:ScalarEquation}
\left(\nabla^\mu-iqA^\mu\right)\left(\nabla_\mu-iqA_\mu\right)\Psi_0=\frac{4\Lambda}{6}\Psi_0.
\end{equation}
In its turn, spin $s=\pm 1/2$ corresponds to a massless fermion field of charge $q$, 
represented by a  Dirac four-spinor $\Psi_{1/2}$
which satisfies the equation~\cite{Liu:2019rbq}:
\begin{equation}\label{eq:DiracEquation}
\gamma^\mu\left(\partial_\mu-\Gamma_\mu-iqA_\mu\right)\Psi_{1/2}=0,
\end{equation}
where 
$\gamma^\mu$ and $\Gamma_\mu$ are the Dirac gamma matrices and the spin connection matrices, respectively (see App.~\ref{sec:app:Dirac}).

Both equations for scalar and fermion fields can be separated by variables using a mode decomposition as
\begin{equation}\label{eq:modesum}
\Psi_{|s|}(\textrm{x})=\displaystyle\int_{\mathbb{R}}d\omega\sum_{\ell=|s|}^{\infty}\sum_{m=-\ell}^{\ell}{}_{|s|}c_{\indmode}\, e^{-i(\omega t-m\varphi)}{}_{|s|}\psi_{\indmode}(r,\theta),
\end{equation}
 where $\textrm{x}$ is a spacetime point,  ${}_{|s|}c_{\indmode}\in \mathbb{C}$  are constant coefficients, $\omega$ is the mode frequency, $m$ is the azimuthal number and $\ell$ is the multipolar number. We can further decompose ${}_0\psi_{\indmode}=R_{0}(r)S_{0}(\theta)$ for the spin-0 scalar and ${}_{1/2}\psi_{\indmode} = (F_- \ G_+ \ G_- \ F_+)^T$, where $F_\pm \equiv \mp R_{-1/2}(r)S_{\pm1/2}(\theta)/[(r\pm ia\cos\theta)\sqrt{2}]$ and $G_\pm \equiv \pm R_{+1/2}(r)S_{\pm1/2}(\theta)$, for the spin-1/2 Dirac spinor. The radial $R_s$ and angular $S_s$ factors satisfy master ordinary differential equations (ODEs), in the sense that the spin $s=0$, $\pm 1/2$ appears as a parameter~\cite{suzuki1998perturbations}:
  \begin{align}
  \label{eq:radialeq}
  \left[
  \Dr^{-s}\del[r]\Dr^{s+1}\del[r]+
  \frac{W^{2}-isW\Dr'}{\Dr}
  + 2isW' -Y
  \right]
  \R=0,
  \\
   \label{eq:angulareq}
   \left[
     \del[u]\Delta_{u}\del[u] -  \frac{1}{\Delta_{u}}\left(
        H+\frac{s}{2}\Delta_{u}'
      \right)^{2} + 2 s H' - X
  \right]
  \Ang =0,
\end{align}
where $u \equiv \cos\theta$,
a prime on a function means derivative with respect to its argument,
with
\begin{align}
  \label{eq:2HW}
  W(r)&\equiv \Xi\,[\omega(r^{2}+a^{2})-am]-qQ r,
  \nonumber
  \\
    Y &\equiv \frac{2}{L^{2}}(s+1)(2s+1)r^{2} + \lamslm,
\end{align}
and
\begin{align}
  H(u) &\equiv \Xi[a\omega(1-u^{2})-m], \ \Delta_u(u) \equiv (1-u^{2})(1+\alpha u^{2}),
 \nonumber \\
 X &\equiv 2(2s^{2}+1)\alpha u^{2} -\lamslm-s(1-\alpha).\,  
 \label{eq:angular-poly}
\end{align}
Here, $\lamslm$ is the separation constant. 
For $\Ang$, we choose \PRDvtwo{boundary} conditions such that the solutions of the angular ODE \eqref{eq:angulareq} are regular at the two endpoints $u=\pm 1$
so the separation constant $\lamslm=\lamslm(\omega)$ becomes the angular eigenvalue. 
 \PRDvtwo{Explicitly, we require that~\cite{yoshida2010quasinormal}
\begin{equation}
S_s(u) \sim \left\{\begin{array}{ll}
     (u+1)^{|m-s|/2},& \textrm{as }u\to-1, \\
     (u-1)^{|m+s|/2},& \textrm{as }u\to1.
\end{array}\right. 
\end{equation}}
The angular eigenvalue $\lamslm$, when evaluated in Kerr (i.e, for $L\to \infty$), is equal to $\lambda$ in Teukolsky's work in Ref.~\cite{Teukolsky:1973ha}. Thus, in particular, $\lamslm$ for $L\to \infty$ reduces to $\ell(\ell+1)-s(s+1)$ when $a=0$. For reference, our definition of $\lamslm$ is related with the eigenvalue $\lambda$ of Refs.~\cite{suzuki1998perturbations,yoshida2010quasinormal} as $\lamslm= \lambda-2s(1-\alpha)$. The eigenvalue $\lamslm$  has the following properties: (i) ${}_{-s}\lambda_{\ell m} = \lamslm +2s(1-\alpha);$ (ii) $\lamslm = {}_{s}\lambda_{\ell,-m}(-\omega)$;
(iii) $\lamslm(\omega^*) =\lamslm^*(\omega)$.

It is easy to show that, near the horizons $r_j$, $j\in\{\indmm,-,+,c\}$, $\R$ behaves as:
\begin{equation}
\R \sim c_1^{(j)} (r-r_j)^{-s}e^{-i\omega_j r_*}+c_2^{(j)} e^{+i\omega_j r_*}, \quad r\to r_j,
\end{equation}
for some complex coefficients $c_1^{(j)}$ and $c_2^{(j)}$, with 
\begin{equation}
\boN{j}\equiv\omega-m\Omega_j-q\phi_j, 
\end{equation}
and where $r_*$ is defined via $dr_*\equiv \Xi(r^2+a^2)dr/\Dr$, choosing the constant of integration so that:
\begin{equation}
r_* = \frac{\ln|r-\rmm|}{2\kappa_{\indmm}}-\frac{\ln|r-r_-|}{2\kappa_-}+\frac{\ln|r-r_+|}{2\kappa_+}-\frac{\ln|r-r_c|}{2\kappa_c}.
\end{equation}

Mode solutions correspond to \PRDvtwo{solutions $R_s$ of the radial ODE \eqref{eq:radialeq} for} complex frequencies $\omega=\oQNM\in \mathbb{C}$ such that, neglecting constant factors,
\begin{equation}
R_s(r)\sim\left\{\begin{array}{ll}
     (r-r_+)^{-s}e^{-i\bo{+}r_*}, & \text{as } r\to r_+,  \\
     e^{+i\bo{c}r_*}, & \text{as } r\to \rc, 
\end{array}\right.
\label{eq:QNM bc}
\end{equation}
where $\bo{j}\equiv \left.\boN{j}\right|_{\omega=\oQNM}\PRDvtwo{=\oQNM-m\Omega_j-q\phi_j}$ with $j\in \{-,+,c\}$. 
That is, mode solutions are purely ingoing into the event horizon and purely outgoing to the \PRDvtwo{c}osmological horizon \PRDvtwo{(as required from physical considerations -- see, e.g., \cite{Teukolsky:1973ha})}. 
The \PRDvtwo{conditions} \eqref{eq:QNM bc} \PRDvtwo{turn} the radial Eq.~\eqref{eq:radialeq} into an eigenvalue problem similar to that of the angular Eq.~\eqref{eq:angulareq}\PRDvtwo{, with the frequencies $\oQNM$ being the eigenvalues in this instance}. For each value of the pair $(\ell,m)$ there may exist a series  of mode solutions, labelled by the overtone index $n=0,1,2,...$ for increasing values of $|\text{Im}(\oQNM)|$.
Quasinormal modes (QNMs) are mode solutions for \PRDvtwo{$\text{Im}(\oQNM)\leq0$} and are the main focus of this paper; unstable modes are mode solutions for \PRDvtwo{$\text{Im}(\oQNM)>0$} (they grow exponentially fast in time).

We note that, because of the so-called Teukolsky-Starobinsky identities~\cite{suzuki1999analytic}, the frequencies of mode solutions for $s=1/2$ are the same as for $s=-1/2$. Therefore, for fermion field perturbations, we focus on $s=1/2$ without loss of generality.

%---------------------------------------------
%---------------------------------------------
\section{QNM families}\label{sec:families}

%----------------------------------------
%\subsection{Neutral field modes}
\subsection{Expressions for the various families of QNM frequencies}
\label{sec:neutral fams}
%----------------------------------------

The QNMs of a KNdS black hole can be divided into four families\footnote{There is in fact an interplay between the various families, which we show in~\cite{Casals:Marinho:prep}.}:
the photon-sphere (PS), the near-extremal (NE), the de Sitter (dS) and the near-Nariai limit (NN)  modes. 

\vspace{0.5cm}
{\it PS family}
\vspace{0.5cm}

The PS modes are associated with the spherical photon orbits of the spacetime~\cite{1972ApJ...172L..95G,PhysRevD.31.290,2009PhRvD..79f4016C,Dolan:2010wr,Yang:2012he,PhysRevD.97.104060,Rahman:2018oso}.
In the eikonal limit, also known as the geometric optics approximation, where  $\ell\gg1$, we can obtain PS modes that are related with the properties of those orbits. In App.~\ref{sec:geo_approx}, we obtain that the PS frequencies of a neutral field in KNdS space-time are
\begin{equation}\label{eq:eikonal}
\oPSn{n} \equiv E_\ph -i\left(n+\frac{1}{2}\right)\LyapLambda,
\end{equation}
where $\LyapLambda$ is the Lyapunov exponent (i.e., the inverse of the instability timescale) of the co-rotating equatorial photon orbit and $E_\ph$ is related to the energy of this orbit.
For low values of $n$, the real and imaginary parts of Eq.~\eqref{eq:eikonal} agree well with our numerical results across all values of spacetime parameters $\{M,a,Q,\Lambda\}$ that we considered in this work
when the value of the multipolar number $\ell$ is large and for general values of the azimuthal number $m$.
We have numerically observed that, when the field has charge $q$, a better approximation to the real part is that in \eqref{eq:eikonal} plus $q\, \phi(r_\ph)$. 

\vspace{0.5cm}
{\it NN family}
\vspace{0.5cm}

As stated before, the KNdS spacetime can be extremal in two different ways: by taking $r_+\to r_-$ (called the extremal black hole limit) and by taking $r_+\to r_c$ (called the rotating Nariai limit). Near both of these extremes, we have zero-damped QNMs, i.e., modes with the imaginary part going to zero as extremality is approached. As for the $r_+\to r_c$ limit, it was analytically shown in~\cite{Novaes:2018fry} that, in Kerr-dS, \PRDvtwo{two subfamilies of neutral field modes, one that corresponds to an expansion about $m\Omega_+$ and the other one about $m\Omega_c$, behave as:}%\sout{two families of neutral field modes have the following behaviour: }
\begin{equation}
\omega_{\text{NN}}^{\PRDvtwo{+,c}}=m\Omega_{+,c}-i(r_c-r_+)\tilde{\kappa}\left(n+\frac{1}{2}\right),
\end{equation}
where $\tilde{\kappa}$ is a non-vanishing  factor which only depends on the spacetime parameters. 
In fact, this result also holds for KNdS (for field spin-0 and spin-1/2), since 
 the radial (as well as angular) ODEs for KNdS and Kerr-dS for  $q=0$ are formally the same  when written in terms of the black hole radii (e.g. see Ref.~\cite{suzuki1998perturbations}).

In our search of spacetime parameters for which there could possibly be violation of the linear version of SCC, we effectively look for $\beta>1/2$ (see Sec.~\ref{sec:beta}). 
In the limit where the NN modes emerge (namely, $r_+\to r_c$), clearly $\text{Im}(\omega_{\text{NN}}^{\PRDvtwo{+,c}})\to 0$, so that $-\text{Im}(\omega_{\text{NN}}^{\PRDvtwo{+,c}})/\kappa_-$ is very small, thus ensuring SCC in the near-Nariai limit (we have observed that this behaviour also holds for charged fields). For this reason, we do not show in our plots spacetime regions near the Nariai limit.

\vspace{0.5cm}
{\it NE family}
\vspace{0.5cm}

In their turn, the NE frequencies $\oNEn{n}$ are obtained by carrying out near-extremal black hole asymptotics $r_-\to r_+$ while assuming that their imaginary part goes to zero (e.g.,~\cite{Hod:2017gvn} in Reissner-Nordstr\"om,~\cite{detweiler1980black,yang2013quasinormal} in Kerr and~\cite{Dias:2018ufh,Cardoso:2017soq} in RNdS).
We carried out a  similar near-extremal analysis in KNdS and obtained~\cite{Casals:Marinho:prep}:
\begin{align}
\label{eq:NE}
\oNEn{n}=&\frac{1}{2}\left[m(\Omega_++\Omega_-)+q (\phi_++\phi_-)\right]-\\&i\frac{\kappa_++\kappa_-}{2}\left[n+\frac{1}{2}+\varepsilon\sqrt{\tilde{\lambda}_s-\tilde{A}^2+\left(s+\frac{1}{2}\right)^2}\right],\nonumber
\end{align}
where $\varepsilon=+1$ when the argument of the square root is positive and $\varepsilon=-1$ when it is negative, with
\begin{equation}
\tilde{A}\equiv\lim\limits_{r_-\to r_+}\left(m\frac{\Omega_- -\Omega_+}{2\kappa_+}+q\frac{\phi_- - \phi_+}{2\kappa_+}\right)
\end{equation}
and
\begin{equation}
\tilde{\lambda}_s\equiv \lim\limits_{r_-\to r_+}\frac{L^2\lamslm(m \Omega_+ + q\phi_+)+2r_+^2(1+s)(1+2s)
}{(\rc-r_+)(\rc+3r_+)}.
\end{equation}
\PRDvtwo{Although we give the explicit derivation of Eq.~\eqref{eq:NE} in Ref.~\cite{Casals:Marinho:prep}, let us here briefly comment on how  it is obtained. We first expand the radial Eq.~\eqref{eq:radialeq} for small $(r_+-r_-)/L$ and fixed $(r_+-r)/(r_+-r_-)$. This effectively removes the singularity at $r=r_c$.
Thus, while we continue to use the standard QNM condition at  $r=r_+$, we replace the standard QNM condition at $r=r_c$ by the following condition.
We require that, as $(r_+-r)/(r_+-r_-)\to-\infty$, the radial solution either does not diverge or has an outgoing phase velocity,  depending on whether $\varepsilon=+1$ or $\varepsilon=-1$ respectively (see \cite{Dias:2018ufh} for details in RNdS).
This {\it ad hoc} boundary condition, together with the QNM boundary condition at $r=r_+$, leads to Eq.~\eqref{eq:NE}. 
From this analysis, Eq.~\eqref{eq:NE} is expected to be a better approximation to the NE QNMs the larger the value of $(r_c-r_+)/(r_+-r_-)$. Furthermore, we have checked numerically
that Eq.~\eqref{eq:NE}  is a good approximation in a large region of parameter space near extremality.}

\vspace{0.5cm}
{\it dS family}
\vspace{0.5cm}

Finally, the dS modes are associated with KNdS spacetime  with $M=Q=0$ (i.e., vacuum and without event horizon. We see numerically that the dS frequencies
for a neutral field are well approximated by (see~\cite{LopezOrtega:2006ig} in pure dS and~\cite{Cardoso:2017soq,Destounis_2019} in RNdS)\footnote{We do not claim that Eq.~\eqref{eq:dS} is the actual asymptotics of the KNdS QNMs in the limits  $M, Q\to 0$. Instead, Eq.~\eqref{eq:dS} is just a useful expression as a seed (see Sec.~\ref{sec:num meth}) for our numerical calculation in a certain region of parameter space.}
\begin{equation}\label{eq:dS}
\odSn{n} \equiv m \OdS -i(\ell+n+1)\kdS,
\end{equation}
where 
\begin{align}
\OdS\equiv \frac{a}{a^2+L^2},
%\nonumber\\ 
\quad
\kdS\equiv \frac{L}{a^2+L^2}
\end{align}
are, respectively, the angular velocity and surface gravity of the cosmological horizon $r=\rc$ for $M = Q = 0$. 
When the field has charge $q$, we observed numerically that the real part is approximated by that in \eqref{eq:dS} plus $\left.q\, \phic\right|_{M=Q=0}$.

\subsection{Dominant QNMs}\label{sec:dominant}

Let us  discuss here the dominance at late times
of the various QNMs for neutral fields (we consider the behaviour for nonzero $q$  in the next subsection).

We first consider the competition between the NE and PS families for dominance in the near-extremal black hole limit.
Importantly, there exists a critical value $\acrit$ such that, for $a< \acrit$, the square root in Eq.~\eqref{eq:NE} is real for any $\ell$ and $m$. 
On the other hand, for $a>\acrit$, the square root becomes complex for $m$ beyond some large enough value.
Furthermore, we show in App.~\ref{sec:geo_approx} that, as $r_+\to r_-$, $\LyapLambda\sim \kappa_+ \to 0$ if $a\gtrsim \acrit$ whereas $\LyapLambda$ asymptotes to a nonzero value if $a\lesssim \acrit$.
Consequently, for $a\gtrsim\acrit$, $\text{Im}\left(\oPSn{n}\right)$ goes to zero and $\text{Im}\left(\oPSn{n}\right)/\kappa_-$ goes to a nonzero finite value; we have checked that, in this case, the PS family is the dominant one.
On the other hand, for $a\lesssim \acrit$, $\text{Im}\left(\oPSn{n}\right)$ does not go to zero and the NE family is the dominant one.

\PRDvtwo{In any spacetime region for the dS family and for $a\lesssim \bar{a}$ with $r_+\to r_-$ for the NE family, we have numerically checked that the slowest decaying modes are those with $\ell=m=|s|$($=0$, $1/2$); as expected from Eq.~\eqref{eq:dS} in the dS case. The PS family, for any large but finite $\ell$, is dominated by the $\ell = m$ mode, which we see numerically in KNdS and is proven analytically in Kerr-dS in Ref.~\cite{Tattersall:2018axd}.} \PRDvtwo{H}owever, the smallest value of $|\text{Im}(\oQNM)|$ over the whole PS family will be given by (the $m$-independent imaginary part of) the $\ell\to \infty$ expression in Eq.~\eqref{eq:eikonal}.

Finally, and as said before, the NN modes never dominate in any  of the regions of spacetime that are relevant for SCC.

%----------------------------------------
\subsection{Behaviour for increasing field charge}\label{sec:incr q}

%----------------------------------------

Let us now analyze the QNM frequencies in terms of the field charge.
The symmetries of the ODEs \eqref{eq:radialeq} and  \eqref{eq:angulareq} together with the boundary conditions in \eqref{eq:QNM bc} implies that, under \PRDvtwo{$\{m,q\}\mapsto \{-m,-q\}$}, we have \PRDvtwo{$\oQNM \mapsto -\oQNM^*$}. 
\PRDvtwo{C}onsider\PRDvtwo{, for the rest of this paragraph,} the $m=0$ modes \PRDvtwo{for the scalar field}\PRDvtwo{. F}or $q=0$ they lie symmetrically with respect to the negative imaginary axis, but that symmetry is broken as $|q|$ increases.
This leads to two different behaviours in the PS  modes as $|q|$ increases, depending on \PRDvtwo{the sign of $\text{Re}\left(\oQNM\right)$ }for $q=0$.
The dS and NE modes, as opposed to any other modes, lie on the imaginary axis for $q=0$ and, as $|q|$ increases, move away from the axis without splitting into different modes.

Within the regions of parameter space that we checked, the dominance of modes within each family is as in the neutral $q=0$ case as long as $q$ is not ``too" large (specifically, within the perturbative regime described \PRDvtwo{below). That is,} within the PS family, the larger the value of $\ell=m$ the more  dominant the mode is; within the dS and NE families, $\ell=m=|s|$ is the dominant mode.
We shall see in Sec.~\ref{sec:SCC}, however, that  this dominance rule for NE is not upheld in the non-perturbative regime, where wiggles appear.

A particularly important limit is that of large field charge. We  carried out a We carried out a Wentzel-Kramers-Brillouin (WKB) analysis in that limit for the charged, scalar and fermion fields in KNdS, similar to the analyses in~\cite{Dias:2018ufh,Ge_2019} in RNdS. We next describe our large-$q$ WKB analysis, where we  expand in powers of $q$ and it is understood that in our expressions we  ignore non-perturbative terms.

We start with the following {\it ansatz} for the radial solution\PRDvtwo{, where we factor out the appropriate asymptotic behaviours at the event and cosmological horizons (see \eqref{eq:QNM bc}) and, in the remaining factor, we use a WKB expansion for large $q$}\footnote{The field charge $q$ has the same dimensions as the inverse of a radius. Any expansion for ``large $q$"  in this asymptotic analysis is to be understood as $q$  large compared to any other quantity of the same dimensions. It is also in this sense that $\mathcal{O}\left(q^n\right)$, for $n\in\mathbb{Z}$, is to be understood in this subsection and a similar understanding  applies to $\lamslm=\mathcal{O}(\omega)$.
},
\begin{align}\label{eq:R large-q}
& R_s(r)=
\left(\frac{r}{r_+}-1\right)^{-i\alpha_+-s}\left(1-\frac{r}{r_c}\right)^{-i\alpha_c}e^{-q\psi(r)}
\times\\&
\left(\phi^{(0)}(r)+\frac{\phi^{(1)}(r)}{q}+\mathcal{O}\left(q^{-2}\right)\right), \nonumber
\end{align}
for some functions $\psi(r)$ and $\phi^{(0),(1)}(r)$ and where
\begin{equation}
\alpha_j\equiv \frac{\omega_j}{2\kappa_j}, \quad j=+,c,
\end{equation}
are related to the Frobenius characteristic exponents of the radial ODE \eqref{eq:radialeq} at $r=r_{+,c}$.
Next, we insert \eqref{eq:R large-q} into the radial ODE and expand for large $q$ with  $\omega=\omega^{(1)} q+\omega^{(0)}+\mathcal{O}\left(1/q\right)=\mathcal{O}\left(q\right)$ for some coefficients $\omega^{(0),(1)}$,
and obtain the following first-order ODE for $\psi$:
\begin{equation}\label{eq:psi}
q\frac{d\psi}{dr}=\pm i\left(\Xi\,\omega-q\phi(r)\right)\frac{r^2+a^2}{\Delta}-\frac{i\alpha_+}{r-r_+}+\frac{i\alpha_c}{r_c-r}.
\end{equation}
Taking the asymptotics of this equation as $r\to r_j$, (and assuming that $d\psi/dr$ is regular at $r=r_j$)  leads to two distinct expressions for the leading-order large-$q$ coefficient $\omega^{(1)}$, namely,
$\omega^{(1)}_j=\phi_j$, with  $j=+$ given by the lower sign in \eqref{eq:psi} and $j=c$ given by the upper sign.

In order to find the next-to-leading order coefficient $\omega^{(0)}$, 
we re-insert \eqref{eq:R large-q} into the radial ODE and expand for large $q$ with $\omega= \phi_j q+\omega^{(0)}+\mathcal{O}\left(1/q\right)=\mathcal{O}\left(q\right)$ and $\lamslm=\lambdaq\, q+\mathcal{O}(1)$
for some coefficient $\lambdaq$ (since $\lamslm=\mathcal{O}\left(\omega\right)=\mathcal{O}\left(q\right)$ \PRDvtwo{--}  see~\cite{Casals:Marinho:prep} in K(N)dS and, e.g.,~\cite{Casals:2004zq} for the proof that $\lamslm=\mathcal{O}(\omega)$ as $\omega\to\infty$ in Kerr).
We then pick the leading order in $q$ of the resulting equation and expand it about $r=r_j$.
In this expansion, we replace $q\,d\psi/dr$ by the result of expanding  the right-hand side of \eqref{eq:psi} with $\omega=\phi_jq+\omega^{(0)}+\mathcal{O}\left(1/q\right)$ first for large $q$ and then about $r=r_j$ (for the corresponding  signs in the right-hand side of \eqref{eq:psi}).
By solving the resulting expansion of the radial ODE we finally obtain $\omega^{(0)}$.
The result for our leading-order large-$q$ WKB expression for the QNM frequencies $\bo{j}=\oQNM-m\Omega_j-q\phi_j$, with $j\in \{+,c\}$, is
\begin{equation}\label{eq:QNM large-q}
\oq{j}\equiv
s_j\frac{\kappa_j(r_j^2+a^2)}{2Q(r_j^2-a^2)}
\lambdaq
-\frac{i}{2}\kappa_j,
\end{equation}
 where
$s_+\equiv +1$, $s_c\equiv -1$.
The expression for $j=+$ corresponds to
a ``black hole family" and that for $j=c$ to a ``cosmological family".
These two WKB families $j=+,c$ are the limiting values for the two  different  behaviours in the PS modes mentioned above  when the symmetry is broken as $q$ increases.
Furthermore, the NE modes asymptote to $\oq{+}$ for large-$q$
(as can be checked by comparing \eqref{eq:NE} for large-$q$ with 
\eqref{eq:QNM large-q} for $j=+$ in the near-extremal limit) -- this provides a consistency check of our analytic expressions.

%----------------------------------------------------
%---------------------------------------------------------------------------------------------------------
\section{NUMERICAL METHOD}\label{sec:num meth}

In order to numerically obtain the QNM frequencies \PRDvtwo{(up to an arbitrary precision)}, we extended to KNdS the method that Leaver  \cite{Leaver:1985} originally developed and applied to Kerr spacetime and was later extended to Kerr-dS in Ref.~\cite{yoshida2010quasinormal}. However, apart from extending to KNdS, we also changed the method in~\cite{yoshida2010quasinormal} in a way which we detail -- and justify -- below.
This method is equally valid for any
 mode solution frequencies $\oQNM$, whether QNMs or unstable modes.
 
In a nutshell, our method essentially consists of, on the one hand, carrying out a power series expansion of the angular solution $S(u)$ about one endpoint (namely, $u=-1$ or $u=1$) and then requiring regularity at the other endpoint; this leads to a continued fraction equation for the eigenvalue $\lamslm(\omega)$.
Similarly, we expand the radial solution $R(r)$ essentially \PRDvtwo{(in the precise manner described below)} about the cosmological horizon $r=r_c$, so that it readily  obeys the condition in Eq.~\eqref{eq:QNM bc} as $r\to r_c$. Then, requiring that this expansion satisfies the condition in Eq.~\eqref{eq:QNM bc} as $r\to r_+$ leads to another continued fraction equation (which involves $\lamslm$) for the mode solution frequency $\oQNM$.

Let us see this more explicitly. Let us first carry out a M\"{o}bius transformation as
\begin{equation}
z \equiv \frac{\left(r_- -\rmm\right)}{\left(r_- - r_c\right)}\frac{\left(r - r_{c}\right)}{\left(r -\rmm\right)}. 
\end{equation}
This leads to the map
\begin{equation}
r = (\rmm, r_-,r_{+},r_c,\infty) \mapsto z = (\infty, 1, x,0,\zeta_\infty),
\end{equation}
where 
\begin{equation}
\zeta_\infty \equiv \frac{r_- -\rmm}{r_- - r_c},\ \ x \equiv \zeta_\infty \frac{r_+ - r_{c}}{r_+ -\rmm}.
\end{equation}
We now express the radial function $R_s$ as:
\begin{equation}
\label{eq:R series}
R_s(z) =z^{iB_c}(z-1)^{iB_-}(z-x)^{-s-iB_+} (z-\zeta_\infty)^{2s+1}g(z),
\end{equation}
for some function $g(z)$, where 
\begin{equation}
B_j \equiv \frac{W(r_j)}{\Delta'_r(r_j)}=\frac{\Xi(r_j^2+a^2)}{\Dr'(r_j)}\omega_j,\quad  j=(-,+,c).
\end{equation}
More explictly,
\begin{eqnarray*}
B_{\pm} &=& \Xi\frac{(r_\pm^2+a^2)\omega-am-qQr_\pm/\Xi}{(r_c-r_\pm)(\rmm-r_\pm)(r_\mp -r_\pm)/L^2}, \nonumber \\
B_c &=& \Xi\frac{(r_c^2+a^2)\omega-am-qQr_c/\Xi}{(r_+-r_c)(\rmm-r_c)(r_- -r_c)/L^2}.
\end{eqnarray*}
By inserting expression \eqref{eq:R series} into the radial ODE \eqref{eq:radialeq} 
we find that $g(z)$ satisfies a Heun equation~\cite{ronveaux1995heun}:
\begin{align}\label{eq:radialequation_f(z)}
& 0=g''(z)+\left[\frac{1+s+2iB_c}{z}+\frac{1+s+2iB_-}{z-1}+
\nonumber \right.\\ &\left.
\frac{1-s-2iB_+}{z-x}\right]g'(z)+\frac{b_1b_2 z+v_{x}}{z(z-1)(z-x)}g(z),
\end{align}
with $b_1 \equiv 1+2iB_c+2iB_-$, $b_2\equiv 1+s-2iB_+$ and the accessory parameter
\begin{multline}
v_{x}\equiv i(B_+-B_c)-4B_+ B_c-i(1+2s)(B_c+B_-)x+\\
s(s+1) + \frac{(1+s)(1+2s)(r_c\rmm+r_+r_-)-\lamslm L^2}{(r_c-r_-)(r_+-\rmm)}.
\end{multline}

Let us consider the following \PRDvtwo{Taylor series expansion in $z$ as an} \textit{ansatz} for a solution of Eq.~\eqref{eq:radialequation_f(z)}:
\begin{equation}\label{eq:localHeun_rc_z0}
y(z)=\sum_{\ik=0}^{\infty}d_\ik z^\ik
\end{equation}
for some coefficients $\{d_\ik\}$.
By inserting it into Eq.~\eqref{eq:radialequation_f(z)}, we  obtain the following three-term recurrence relation for these coefficients:
\begin{equation}\label{eq:rec rln}
\an{\ik} d_{\ik+1}+\bn{\ik} d_\ik+\cn{\ik} d_{\ik-1} =0, \qquad \forall\, \ik=0,1,2,\dots
\end{equation}
with $d_0=1$, $d_{-1}=0$ (so that $d_{\ik}$ satisfies the ``boundary condition" $d_{1}/d_{0}=-\bn{0}/\an{0}$) and 
\begin{align}\label{eq:a,b,c}
\an{\ik} &\equiv  (\ik+1)(\ik+1+s+2iB_c)x, \nonumber\\
\bn{\ik} &\equiv v_x -\ik^2(1+x)-\ik(b_2+xb_1-s+2sx+2iB_c),\nonumber\\
\cn{\ik} &\equiv \left[\ik+2i(B_c+B_-)\right](\ik+s-2iB_+).
\end{align}

It is easy to show (see~\cite{baber_hasse_1935}) that the solutions $d_\ik$ to the recurrence relation \eqref{eq:rec rln} have, if $d_\ik\not\equiv 0$, either of the two following linearly independent behaviours for large-$\ik$:
\begin{align}\label{eq:large-n dom}
\lim\limits_{\ik\to\infty}\frac{d_{\ik+1}}{d_\ik}= \frac{1}{x}- \frac{1+s+2iB_+}{\ik x}+\mathcal{O}(\ik^{-2}),
\end{align}
or
\begin{align}\label{eq:large-n min}
 \lim\limits_{\ik\to\infty} \frac{d_{\ik+1}}{d_\ik} =1-\frac{1-s-2iB_-}{\ik}+\mathcal{O}(\ik^{-2}).
\end{align}
If the behaviour is as in \eqref{eq:large-n dom}, then we denote the solution by $d_{\ik}^{(2)}$,
and if it is  as in \eqref{eq:large-n min}, then we denote it by $d_{\ik}^{(1)}$.
Given that $x\in (0,1)$, this implies (see below Theorem 2.2 of Ref.~\cite{gautschi1967computational}) that $d_\ik^{(1)}$ is the (unique) minimal solution and $d_\ik^{(2)}$ is a dominant solution of the recurrence relation Eq.~\eqref{eq:rec rln}, i.e, $\lim\limits_{\ik\to\infty}d_{\ik}^{(1)}/d_\ik^{(2)}=0$. 
Furthermore, the infinite series in \eqref{eq:localHeun_rc_z0} is guaranteed to converge $\forall z\in [0,x)$ if the large-$\ik$ limit \eqref{eq:large-n dom}  is satisfied and $\forall z\in [0,1)$ if, instead, \eqref{eq:large-n min} is satisfied\footnote{Indeed, the expansion in Eq.~\eqref{eq:localHeun_rc_z0} represents a local Heun function~\cite{Bateman:Vol3,ronveaux1995heun} around $z=0$, which is guaranteed to converge on the disk $|z|<x$
but whose convergence at $z=x$ 
 is not guaranteed.}. Therefore, $y$ is always holomorphic $\forall z\in [0,x)$ and, in the specific case that a minimal solution $d_{\ik}$ exists and $d_{\ik}=d_{\ik}^{(1)}$, then $y$ is  holomorphic in the larger interval $z\in [0,1)$.

It is \PRDvtwo{straightforward} to see that the expression for $R_s(z)$ in Eq.~\eqref{eq:R series} has been built so that it satisfies the  \PRDvtwo{boundary conditions in Eq.~\eqref{eq:QNM bc}} if and only \PRDvtwo{if} $g(z)$ is holomorphic in $z\in [0,x]$\footnote{ \PRDvtwo{In its turn, the factor $(z-\zeta_\infty)^{2s+1}$ in Eq.~\eqref{eq:R series} is used in order to remove the singularity of the radial ODE at $r=\infty$.}}. We shall next show that %the \sout{mode solution condition} \PRDvtwo{boundary conditions} Eq.~\eqref{eq:QNM bc}  \sout{is}\PRDvtwo{are} 
\PRDvtwo{there  being a  solution $R_s(z)\not\equiv 0$ satisfying the boundary conditions in Eq.~\eqref{eq:QNM bc}}
is equivalent to a certain continued fraction equation being satisfied; \PRDvtwo{such continued fraction} is the equation that we shall solve in practise in order to calculate frequencies of mode solutions.

{\bf Proposition.}
\PRDvtwo{There exists a nontrivial solution of the radial ODE \eqref{eq:radialeq} satisfying the boundary conditions in Eq.~\eqref{eq:QNM bc} 
if and only if the following continued fraction equation is satisfied}:
\begin{equation}\label{eq:cont frac}
\PRDvtwo{0=\frac{\bn{0}}{\an{0}}}-\dfrac{\cn{1}}{\bn{1}-\dfrac{\an{1}\cn{2}}{\bn{2}-\dfrac{\an{2}\cn{3}}{\bn{3}-\dots}}}
\end{equation}
where $\an{\ik}$, $\bn{\ik}$ and $\cn{\ik}$ are given in Eq.~\eqref{eq:a,b,c}.

\begin{proof}
First, assume that \PRDvtwo{$R_s$} satisfies \PRDvtwo{ Eq.~\eqref{eq:radialeq} and }the \PRDvtwo{boundary conditions Eq.~\eqref{eq:QNM bc}}, so that
 $g\not\equiv 0$\PRDvtwo{, given via \eqref{eq:R series},} is holomorphic in $z\in [0,x]$. 
 Since $y$ in \eqref{eq:localHeun_rc_z0} is holomorphic in $z\in [0,x)$,
by unicity of holomorphic functions, we  have that
$g= y$ (up to a constant factor) in  $ z\in [0,x)$.
This implies that $d_\ik$ is nontrivial (i.e., $d_\ik\not \equiv 0$). This, combined with the large-$\ik$ limits \eqref{eq:large-n dom} and \eqref{eq:large-n min},  implies that the recurrence relation \eqref{eq:rec rln} possesses a minimal solution  $d_{\ik}^{(1)}$. Theorem 1.1 in~\cite{gautschi1967computational} then guarantees that the continued fraction in \eqref{eq:cont frac} converges. \PRDvtwo{This} theorem also shows that, in this case, it is
\begin{equation}\label{eq:ratio min}
\frac{d_{\ik}^{(1)}}{d_{\ik-1}^{(1)}}=
-\dfrac{\cn{\ik}}{\bn{\ik}-\dfrac{\an{\ik}\cn{\ik+1}}{\bn{\ik+1}-\dfrac{\an{\ik+1}\cn{\ik+2}}{\bn{\ik+2}-\dots}}}, \quad \ik\geq 1.
\end{equation}
By choosing $\ik=1$ in \eqref{eq:ratio min} and equating it to the boundary condition $d_{1}^{(1)}/d_{0}^{(1)}=-\bn{0}/\an{0}$, Eq.~\eqref{eq:cont frac} follows. 

Let us now turn to the \PRDvtwo{proof of the} implication in the opposite direction. 
Assume that  \eqref{eq:cont frac}  is satisfied. This means, in particular, that the continued fraction in \eqref{eq:cont frac} converges. Because of the latter,
Th.1.1 in~\cite{gautschi1967computational} guarantees that there exists a sequence $d_\ik^{(1)}\not\equiv 0$ which: (i)  satisfies \eqref{eq:ratio min}, and (ii)
is a minimal solution of \eqref{eq:rec rln}. Point (ii) implies that its large-$\ik$ behaviour is as in \eqref{eq:large-n min}.
It then follows from the ratio test that the infinite series in \eqref{eq:localHeun_rc_z0}, with $d_\ik=d_\ik^{(1)}$, converges in $z\in [0,1)$.
Thus, $y$ with $d_\ik=d_\ik^{(1)}$ is  holomorphic in $z\in [0,x]$. Furthermore, point (i) (i.e., \eqref{eq:ratio min} being satisfied) together with  \eqref{eq:cont frac} imply that
the boundary condition $d_{1}^{(1)}/d_{0}^{(1)}=-\bn{0}/\an{0}$ is satisfied. 
Thus,  by choosing $g\equiv y\not\equiv 0$ in $z\in [0,x]$ we have that \PRDvtwo{$R_s$, given by \eqref{eq:R series},} is a mode solution.
This completes the proof.
\end{proof}
A similar continued fraction for the eigenvalue $\lamslm$ may be derived from the angular ODE and it is already given explicitly in Eq.~(38) in Ref.~\cite{yoshida2010quasinormal}\footnote{There is a typographical error in Eq.~(28) in Ref.~\cite{yoshida2010quasinormal}: there should be a minus sign in the definition of $z_s$.} (with coefficients given in Eqs.~(31)-(33) therein)\PRDvtwo{.} Thus, there are two continued fraction equations to be solved:  a radial one for the mode solution frequency  $\oQNM$ and an angular one for the eigenvalue $\lamslm$. 

%When solving for either continued fraction, we always required at least 12 digits of precision.
In order to solve these continued fractions, we use the numerical algorithm of the software \emph{Mathematica} to find the roots of  equations and, for that, we need to provide seeds. For the angular eigenvalues, Ref.~\cite{suzuki1998perturbations} has given an expansion for $\lamslm$ for $|a\omega|\ll1$ and $|a/L|\ll1$\PRDvtwo{, to which we shall loosely refer as a ``small-$a$" expansion}. Thus, if we choose $a$ sufficiently small, we can use that expansion as a seed for the numerical calculation of $\lamslm$. In order to find the eigenvalue for a given value of $a=a_\text{M}$ (not necessarily small), we calculate the eigenvalues for $a=a_k\equiv k\, a_\text{M}/N_\lambda$, $k=1,2,\dots,N_\lambda$, for some $N_\lambda\in\mathbb{N}$, as follows. We use the small-$a$ expansion of Ref.~\cite{suzuki1998perturbations} as the starting point to find the ``exact'' eigenvalue for the initial $a=a_1$. We then proceed by increasing $k$ by $1$ at a time: for finding the eigenvalue for $a=a_k$, $k=2,3,\dots,N_\lambda$, we use the previously obtained eigenvalue for $a=a_{k-1}$ as a seed; we repeat this process $N_\lambda-1$ times until obtaining the eigenvalue for $a=a_{N_\lambda}=a_\text{M}$. In this way, we can track the eigenvalue for a given $(\ell,m)$-mode as $a$ increases.

In order to search for unstable modes, we plotted the r.h.s.~of Eq.~\eqref{eq:cont frac} in regions of the upper complex-frequency plane and looked for  \PRDvtwo{its} zeros.
We  report here that we found no such zeros, and so no unstable modes, for the values of the KNdS spacetime parameters which we consider in this paper.
We thus henceforth focus on QNMs.

For computing QNM frequencies we solved the coupled set of the two continued fraction equations for the eigenvalue $\lamslm$ and the QNM frequency  $\oQNM$.

The seeds for the QNM frequencies $\oQNM$ are trickier to find than those for the \PRDvtwo{angular} eigenvalue $\lamslm$. A computationally expensive \PRDvtwo{approach for} finding \PRDvtwo{seeds for the QNM frequencies}  is by plotting the r.h.s.~of Eq.~\eqref{eq:cont frac} in a region of the lower complex-frequency plane and seeing roughly where its roots are (similarly to the search for unstable modes in the upper plane), so as to use them as seeds for the numerical routine.  
\PRDvtwo{An alternative, faster approach for obtaining seeds for the QNM frequencies is by directly using the analytical expressions in Eqs.~\eqref{eq:eikonal}, \eqref{eq:NE} and \eqref{eq:dS} when these expressions yield a good approximation for QNMs.
We combined these two approaches in order to make sure that we did not miss any QNM frequency $\oQNM$ which was relevant for SCC for some initial set of black hole and field parameters which we denote by $\mathcal{P}_1$.
We then proceeded similarly to the way we did for the angular eigenvalue $\lamslm$.
Namely, we numerically obtained QNM frequencies for a set, say $\mathcal{P}_{2}$, of values of black hole and field parameters slightly different from those in $\mathcal{P}_1$ by using as seeds the values of the frequencies that we previously numerically obtained for $\mathcal{P}_{1}$.
We then again changed to a new set $\mathcal{P}_{3}$ of slightly different values of black hole and field parameters and used as seeds the numerical frequencies obtained for $\mathcal{P}_{2}$. And we proceed in a similar manner with new sets $\mathcal{P}_{4}, \mathcal{P}_{5},\dots$ until covering the desired region of parameter space.}

We finish this section with some comments on our method and checks that we implemented. The expression for the radial solution $R_s$ that we obtain from \eqref{eq:R series} together with the expansion for $g(z)=y(z)$ as in Eq.~\eqref{eq:localHeun_rc_z0} is similar to the expression in Eq.~(50) in Ref.~\cite{yoshida2010quasinormal} (in Kerr-dS) with an important difference: here $y(z)$ is essentially expanded about the cosmological horizon whereas in Ref.~\cite{yoshida2010quasinormal} it is about the event horizon. The reason for this change is the following. When we numerically implemented a routine for finding zeros of the continued fraction given by Eqs.~(52)-(55) of \cite{yoshida2010quasinormal}, we only managed to reproduce the QNM tables  and plots in \cite{yoshida2010quasinormal} when $|x_r|<1$ (where $x_r$ is defined in Eq.~(49) of the same article) but not when $|x_r|>1$ (even though they state that they applied their method for $|x_r|>1$). On the other hand, we managed to reproduce all QNM data provided in \cite{yoshida2010quasinormal}, for any $x_r$, using an implementation of our continued fraction \eqref{eq:cont frac}, i.e., using the expansion in Eq.~\eqref{eq:localHeun_rc_z0} for $g(z)$.
This provides a good check of our method and code (at least in Kerr-dS).

As further checks of our method, we found, in KNdS, agreement to all digits with the QNMs in tables in Ref.~\cite{Chang2005} for $|s|=1/2$ as well as visual agreement with Fig.~3 in Ref.~\cite{Konoplya:2007zx} \PRDvtwo{(where they use a  transformation of the radial function and a Taylor series expansion which are different from those here as well as those in \cite{yoshida2010quasinormal})} for $s=0$ and $|s|=1/2$; see App.~\ref{sec:KdS} for further checks in Kerr-dS (i.e., $Q=0$).

%=-------------------------------------------------------
%=-------------------------------------------------------
\section{\texorpdfstring{$\beta$}{beta} CONDITIONS FOR SCC}\label{sec:beta}

In order to investigate the validity of the linear version of SCC, we need to assess the \PRDvtwo{strength} of the matter field perturbations  on the 
right Cauchy horizon \PRDvtwo{$\mathcal{CH}_\text{R}^+$ (recall Fig.~\ref{fig:CP-KNdS})}.
To be precise, the requirement for validity is that weak solutions of the perturbed Einstein equations \PRDvtwo{do not} exist %\marc{is it not the opposite - that validity of SCC means that *no* such solutions exist?}\cassio{you are right. It should be "weak solutions do not exist [...]"}
across the Cauchy horizon and, therefore, for the stress-energy tensor for the matter field, which is to be placed on the right hand side of Einstein's equations as sourcing higher-order metric perturbations, to \PRDvtwo{not} be locally integrable (see, e.g., Ref.~\cite{PhysRevD.97.104060}). 
We next derive the relationship between this requirement and values of $\beta\equiv \alpha_g/\kappa_-$, where $\alpha_g$ is the so-called spectral gap, i.e., the infimum (smallest value) of $-\text{Im}(\oQNM)\geq 0$ over {\it all} QNMs.
We also derive the relationship between the requirement of curvature blowup and values of $\beta$.
Our derivation is an extension to our setup in KNdS of the corresponding derivation in Ref.~\cite{PhysRevD.97.104060} for a spin-0 field in Kerr-dS and in Ref.~\cite{Liu:2019rbq} for a spin-1/2 in RNdS.

We denote by  $\psi_s(\textrm{x})\equiv
e^{-i\omega t +i m \varphi}{}_{|s|}\psi_{\indmode}(r,\theta)$ the field modes including all coordinates -- see Eq.~\eqref{eq:modesum}. Henceforth we use outgoing coordinates $\{u,r,\theta,\phio\}$, which are regular on the (right) Cauchy horizon and are related to the coordinates $\{t,r,\theta,\varphi\}$ of Eq.~\eqref{eq:KNdSmetric} via:
\begin{equation}\label{eq:outgoing-transf}
dt = du +dr_*, \ \ d\varphi = d\phio + dr_\phi\equiv d\phio+\frac{a \Xi}{\Delta_r}dr.
\end{equation}
Following Ref.~\cite{PhysRevD.98.104007}, we carry out a gauge transformation $\boldsymbol{A}\to \boldsymbol{A}+d\chi$ and $\psi_s\to e^{i q\chi}\psi_s$, with $d\chi \equiv Qr\, dr/\Delta_r$ and $\boldsymbol{A}$ is the KNdS electromagnetic potential given in Eq.~\eqref{eq:eletromag_pot}. \PRDvtwo{This gauge transformation is carried out so that it yields electromagnetic potential components $A_\mu$ in outgoing coordinates which are regular at all spacetime horizons.} The modes in this new gauge are $\psi_0(u,r,\theta,\phio)= e^{-i\omega u +i m \phio}\tilde{R}_0(r)S_0(\theta)$ for the scalar field and $\psi_{1/2}(u,r,\theta,\phio)=e^{-i\omega u +i m \phio}(\tilde{F}_- \ \tilde{G}_+ \ \tilde{G}_- \ \tilde{F}_+)^T$ for the fermion field, where $\tilde{F}_\pm \equiv \mp \tilde{R}_{-1/2}(r)S_{\pm1/2}(\theta)/[(r\pm ia\cos\theta)\sqrt{2}]$ and $\tilde{G}_\pm \equiv \pm \tilde{R}_{+1/2}(r)S_{\pm1/2}(\theta)$ (see App.~\ref{sec:app:Dirac} for details of the derivation), where $\tilde{R}_s$ is the corresponding radial factor in outgoing coordinates: $R_s = e^{-iq\chi} e^{i\omega r_*} e^{-im r_\phi} \tilde{R}_s$. It is trivial to show using the Frobenius method that two linearly independent  behaviours of $\tilde{R}_s$ as $r\to r_-$ are:
\begin{equation}\label{eq:psi2}
\tilde{R}_s^{(1)}\equiv g^{(1)}(r),
\quad
\tilde{R}_s^{(2)}\equiv(r-r_-)^{-s+i\omega_-/\kappa_-}g^{(2)}(r),
\end{equation}
where $g^{(1),(2)}(r)$ are smooth and non-vanishing functions at $r=r_-$. 

\subsection{Scalar field}

The stress-energy tensor of a massless conformally-coupled scalar field $\Psi_0$ with charge $q$ is given by
\begin{multline}
T_{\mu\nu}=\frac{1}{6}\text{Re}\biggl\{4(D_\mu\Psi_0)^*D_\nu\Psi_0-g_{\mu\nu}(D_\rho\Psi_0)^*D^\rho\Psi_0\\-2\Psi_0^*D_\mu D_\nu\Psi_0+\left(R_{\mu\nu}-\frac{2\Lambda}{3}g_{\mu\nu}\right)|\Psi_0|^2\biggl\},
\end{multline}
where $D_\mu\equiv \nabla_\mu-iq A_\mu$ and $R_{\mu\nu}$ is the Ricci tensor. Thus, the various terms in $T_{\mu\nu}$ possess one of the following scalar field factors: $\partial_\mu \Psi_0^*\partial_\nu \Psi_0$, $\Psi_0^*\partial_\mu \Psi_0$, $\Psi_0^*\partial_\mu\partial_\nu \Psi_0$ and $|\Psi_0|^2$.  One can verify that the dominant terms as the Cauchy horizon is approached are the ones involving two radial derivatives of the field, i.e., terms with either $\text{Re}(\Psi_0^*\partial_r^2\Psi_0)$ or  $\text{Re}(\partial_r\Psi_0^*\partial_r\Psi_0)$, where both terms go like $(r-r_-)^{-2\text{Im}(\omega)/\kappa_--2}$ as $r\to r_-$, due to Eq.~\eqref{eq:psi2}. The term $(r-r_-)^{-2\text{Im}(\omega)/\kappa_--2}$ is locally integrable when $-\mbox{Im}(\omega)/\kappa_->1/2$. Therefore, violation of the linear version of SCC is equivalent to the derivatives of the real part of the scalar field being locally square integrable, which is equivalent to $\beta>1/2$.

Similarly, boundedness of the curvature corresponds to the boundedness of the stress-energy tensor, which itself corresponds to $\beta\PRDvtwo{\geq}1$ due to Eq.~\eqref{eq:psi2}.

Alternatively, one may view the scalar field as an analog of linear gravitational perturbations (see, e.g., Ref.~\cite{Dafermos:2018tha}). Within this viewpoint, the condition for violation of SCC continues to be $\beta>1/2$, since first-order derivatives of the scalar field are analogs of the Christoffel symbols and these are required to be locally square integrable (e.g., Ref.~\cite{Costa2018}).

\subsection{Fermion field}

The stress-energy tensor of a massless fermion field of charge $q$, represented by a Dirac four-spinor $\Psi_{1/2}$, is given by  (see App.~\ref{sec:app:Dirac} for  definitions of symbols):
\begin{equation}
T_{\mu\nu}=\frac{i}{2}\left[\bar\Psi_{1/2}\gamma_{(\mu}D_{\nu)}\Psi_{1/2}-\PRDvtwo{(}D_{(\mu}\bar\Psi_{1/2}\PRDvtwo{)}\gamma_{\nu)}\Psi_{1/2}\right].
\end{equation}

In the outgoing coordinates, we show in App.~\ref{sec:app:Dirac} that both $\gamma_\mu$ and $\Gamma_\mu$ are regular on the Cauchy horizon for a suitable choice of base vectors. This regularity helps us conclude that the local integrability/boundedness of $T_{\mu\nu}$ at the Cauchy horizon  is guaranteed by
the integrability/boundedness of the factors $|\tilde{R}_{\pm1/2}|^2$, $\text{Re}(\tilde{R}_{-1/2}\tilde{R}_{+1/2}^*)$, $\text{Re}(\tilde{R}_{-1/2}\partial_\mu\tilde{R}_{\pm1/2}^*)$ and $\text{Re}(\tilde{R}_{+1/2}\partial_\mu\tilde{R}_{\pm1/2}^*)$. Among these factors, the dominant contribution will come from   $\text{Re}(\tilde{R}_{+1/2}\partial_r\tilde{R}_{+1/2}^*)$,
that is, by a linear combination of $(r-r_-)^{-2-2\textrm{Im}(\omega_-)/\kappa_-}$ and $(r-r_-)^{-3/2-\textrm{Im}(\omega_-)/\kappa_-}$. The local integrability of both of these terms  over all QNMs corresponds to $\beta>1/2$, just like for the scalar field. The boundedness of the stress-energy tensor corresponds to $\beta\PRDvtwo{\geq}3/2$, by contrast with $\beta\PRDvtwo{\geq}1$ in the case of the scalar field.

%----------------------------------------------------
%---------------------------------------------------------------------------------------------------------

\section{Results for Cosmic Censorship}\label{sec:SCC}

The  value of $\beta$ for a  field perturbation in KNdS is equal to $\inf\{\beta_{\text{PS}},\beta_{\text{NE}},\beta_{\text{dS}}\}$,
where $\beta_{\text{PS/NE/dS}}$ denotes the \PRDvtwo{infimum of $-\text{Im}(\oQNM)/\kappa_-$ over the QNMs of the PS/NE/dS family}  only\footnote{We remind the reader that, as explained in Sec.~\ref{sec:families}\PRDvtwo{,} the NN modes are not dominant in the near-extremal regions considered here.}. The numerical method that we used for calculating the QNM frequencies was described in Sec.~\ref{sec:num meth}. \PRDvtwo{Our values for $\beta$ are a result of a  thorough analysis, in which we were particularly careful to make sure that we did not miss any mode which could potentially invalidate our values for $\beta$.} We next present our results for $\beta$ and SCC, first considering neutral fields and later charged fields.

\vspace{0.5cm}
{\it Neutral fields}
\vspace{0.5cm}

In the following analysis for neutral fields, we used the eikonal approximation given in Eq.~\eqref{eq:eikonal} to obtain $\beta_\text{PS}$ as justified in Sec.\ref{sec:dominant}.

Fig.~\ref{fig:cont,s=0} shows contour plots of $\beta$ for spin-$0$ and spin-$1/2$ neutral fields as a function of $a$ and $Q$ (near $Q_{\text{max}}$) for fixed  $\Lambda$. It clearly shows that there exist regions of spacetime parameter space where $\beta>1/2$ for both spin-fields, 
including subregions where $\beta>1$ for spin-0 (top) and where $\beta>3/2$ for spin-1/2 (bottom). It is worth mentioning that we find $\beta>1$ for spin-$0$ even \PRDvtwo{in the RNdS limit of $a\to0$. This tallies with the fact that $\beta>1$ is possible for sufficiently  massive, {\it minimally}-coupled scalar fields in RNdS~\cite{PhysRevD.98.104007} and that 
including a positive coupling (such as in our conformally-coupled case) in the scalar field equation in KNdS
is effectively equivalent to adding a mass to the field.}
%\sout{for $a\to0$ because we consider a conformally coupled scalar field, which, in terms of the field equation, is effectively equivalent to adding a mass to the field, and sufficiently  massive, minimally-coupled scalar fields in RNdS have  $\beta>1$~\cite{PhysRevD.98.104007}.}
We also note that we consistently found  $\beta<1/2$ for spacetime parameter values outside the ranges in Fig.~\ref{fig:cont,s=0}.

As a general behaviour, we find that  when increasing $a/M$ for fixed $Q$, $\beta$ first decreases and then increases (while not surpassing $\beta=1/2$); the closer $Q$ is to $Q_{\text{max}}$ for fixed $a/M$, the larger the $\beta$.
Thus, the values of $\beta$ larger than $1/2$ occur near extremality.

Fig.~\ref{fig:cont,s=0} also shows the regions where each family dominates. 
For $s=0$ the NE dominates in the region of smaller rotation and larger charge; the dS dominates for both smaller  rotation and charge; and the PS dominates in the rest of the range in the plot. It is similar for $s=1/2$, except that the PS also dominates in the region of both smaller  rotation and charge.

\begin{figure}[!ht]
  \includegraphics[width=0.483\textwidth]{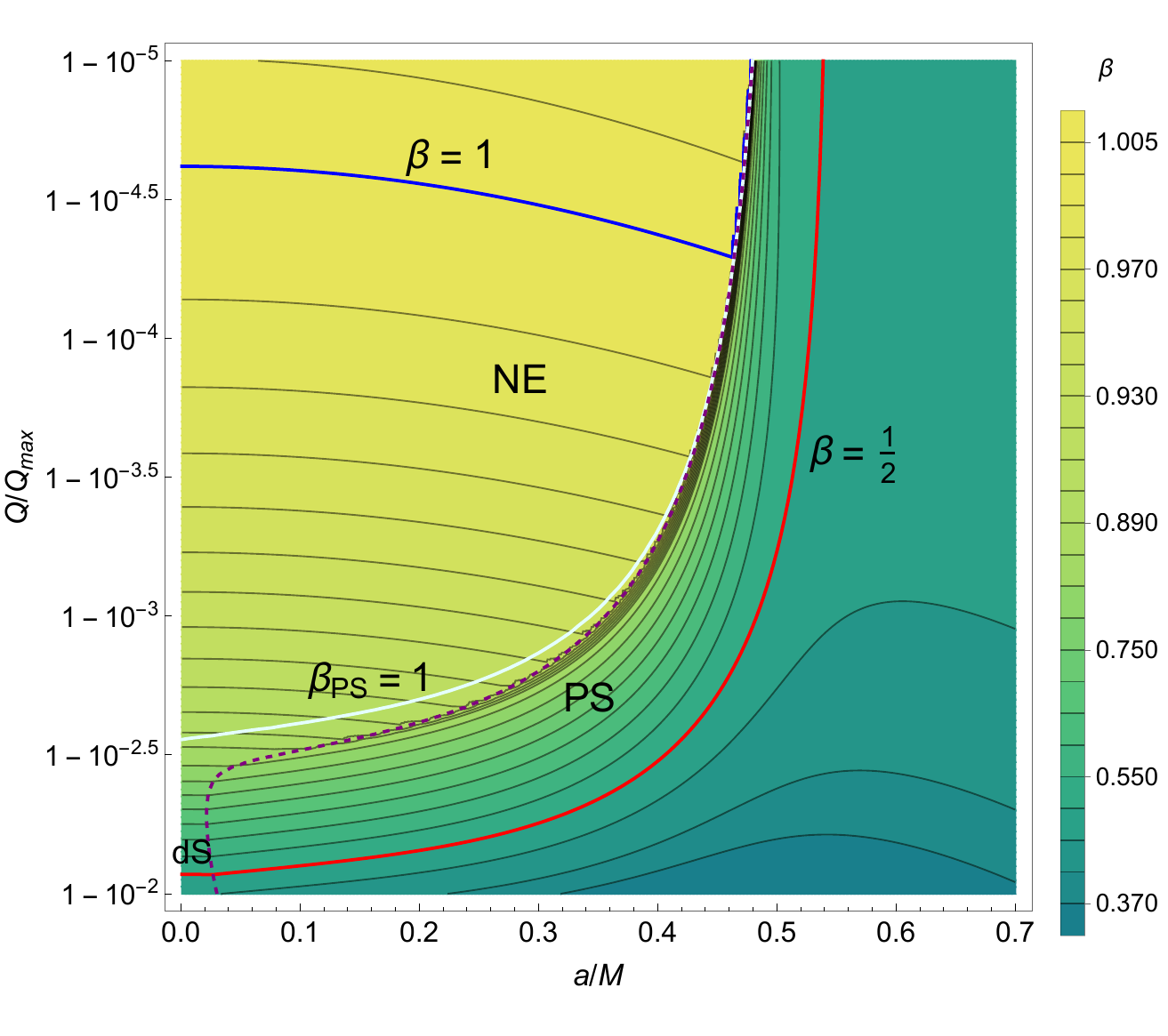}   
  \\
    \includegraphics[width=0.483\textwidth]{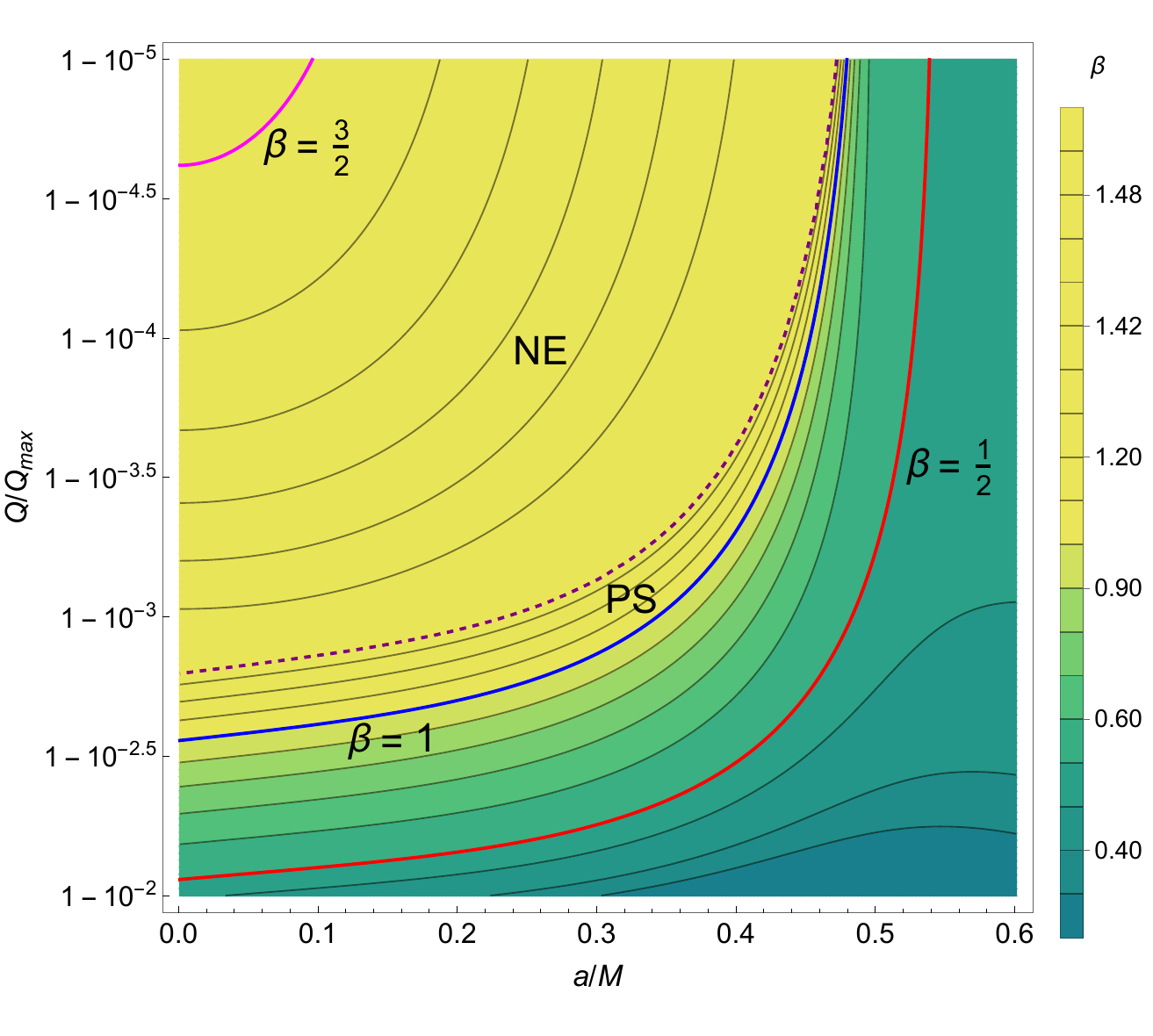}   
    \caption{
Contour plots of $\beta$ \PRDvtwo{(with its scale in yellow-green shades provided on the right of each contour plot)} as a function of $a/M$ and $Q/Q_\text{max}$ for $q=0$, $\Lambda M^2=0.02$; field spin $s=0$ (top) and $s=1/2$ (bottom).
The \PRDvtwo{red}, \PRDvtwo{blue}, and magenta curves correspond to, respectively,  $\beta=1/2$,  $\beta=1$ and $\beta=3/2$ for that spin-field. The \PRDvtwo{light gray} curve for $s=0$ corresponds to $\beta_{\text{PS}}=1$. The \PRDvtwo{purple dashed} curves separate the regions of dominance of the PS, NE and dS families. }
\label{fig:cont,s=0} 
\end{figure}

%-------------------- End of contour plots -------------------------------------------------------------------------------------

\vspace{0.5cm}
{\it As a function of the field charge}
\vspace{0.5cm}

We now look at the effects as we turn on the field charge $q$. It follows from Eq.~\eqref{eq:QNM large-q} that the $\beta$ values for the two large-$q$ WKB families $j\in \{+,c\}$ asymptote to $\beta_j\equiv \kappa_j/(2\kappa_-)$.
Since $\beta_c\to \infty$ and $\beta_+\to \left(1/2\right)^-$ in the extremal limit,  it is the black hole WKB family (i.e., the latter) near extremality that matters for SCC.
Furthermore, since $\beta_+\leq 1/2$, in principle, $\beta$ should not be larger than $1/2$ for large $q$. However, the WKB analysis leading to \eqref{eq:QNM large-q} has ignored non-perturbative terms and so we proceed to show results of our exact numerical investigation.

In the following analysis for charged fields,  we use $\ell=m=10$ [resp. $\ell=m=21/2$] for the  PS family of scalar [resp. fermion] field modes: the true value of $\beta_\text{PS}$ would really be given by $\ell=m\to\infty$ (see Sec.~\ref{sec:incr q}) but we find  (by comparing with the values of $-\text{Im}(\oQNM)/\kappa_-$  provided by PS modes with  $\ell=m$ larger than our mentioned choice) that the value of $-\text{Im}(\oQNM)/\kappa_-$ as provided by the  $\ell=m=10$ [resp. $\ell=m=21/2$] mode is already a sufficiently good approximation.

\begin{figure}[!ht]
    \centering
      \includegraphics[width=0.483\textwidth]{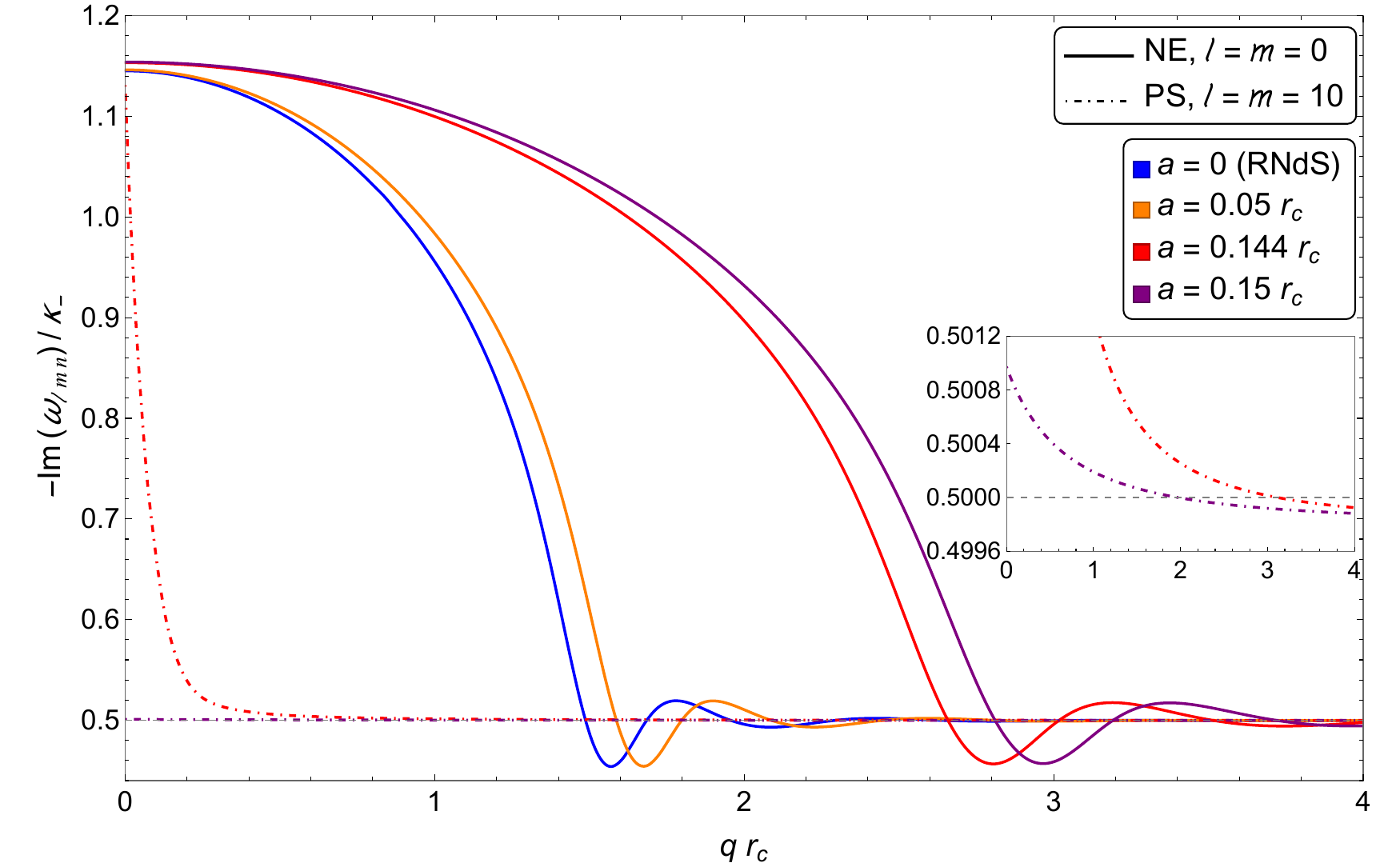}
    \includegraphics[width=0.483\textwidth]{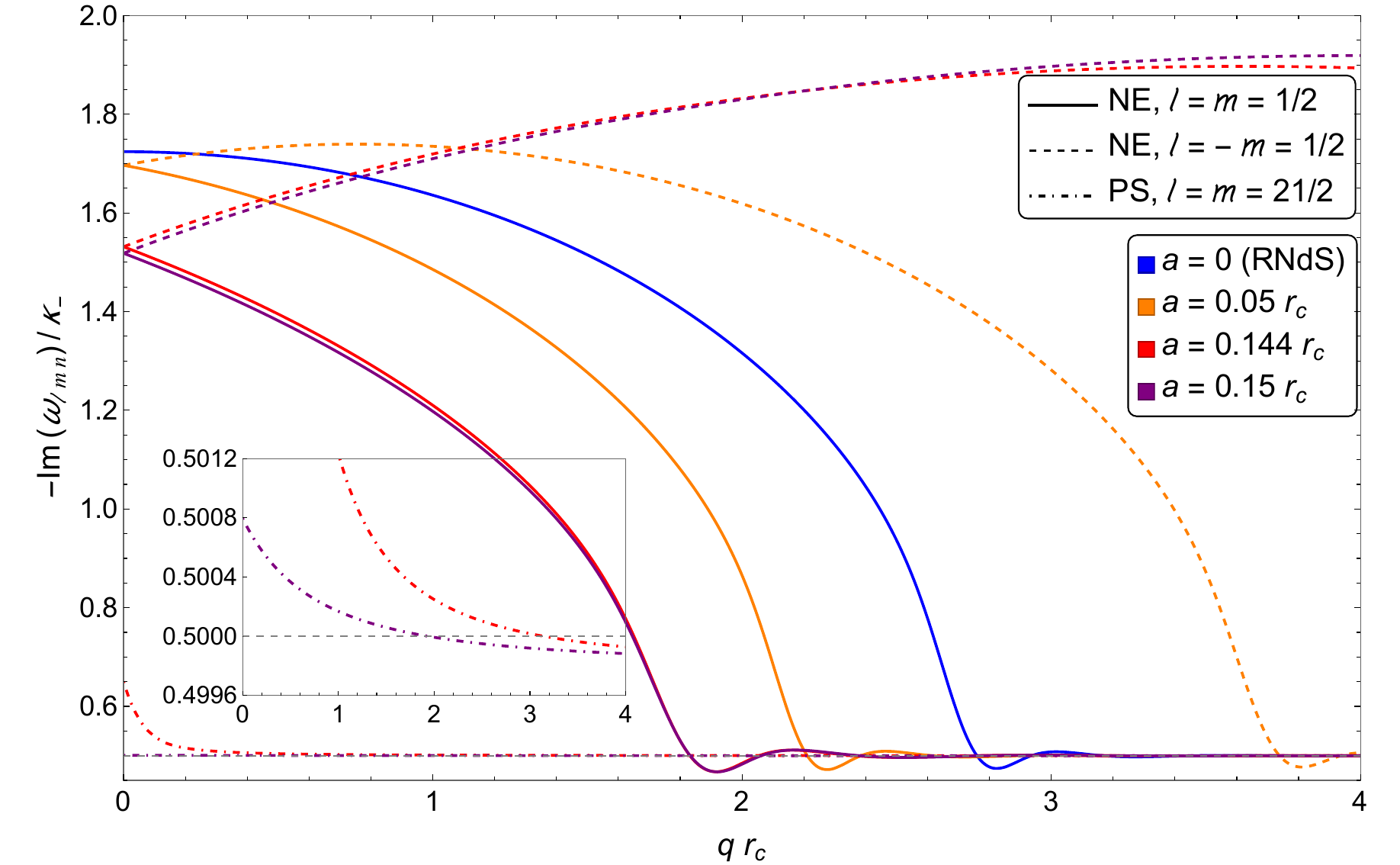}
    \caption{Plots of $-\text{Im}\left(\oQNM\right)/\kappa_-$ as a function of $q\, \rc$ for the most dominant modes for $r_+=\rc/3$ and $Q=\left(1-10^{-4}\right)Q_{\text{max}}$  for various values of $a$ and for $s=0$ (top) and $s=1/2$ (bottom). Here it is $\acrit \approx 0.135\rc$. The $(\ell,m)$ values of the modes used for each family are indicated in the labels; the insets zoom in on the PS family for small $q\, \rc$.}
    \label{fig:wiggles func q}
\end{figure}

Fig.~\ref{fig:wiggles func q} shows the behaviour of the most dominant  QNM frequencies in a region of $q$ for $s=0$ (top) and $s=1/2$ (bottom) near extremality. 
The behaviour of the PS modes as $q$ increases is rather distinct in two regimes.
For $a\lesssim \acritq$ (i.e., angular momentum below the critical value introduced in Sec.~\ref{sec:neutral fams}), 
$\beta_{\text{PS}}$ is well above $1/2$ (and so these do not appear in the plots), at least over a significantly large region of values of $q$,
and so the NE family of modes dominates there. For $a\gtrsim \acritq$, on the other hand, $\beta_\text{PS}$ rapidly approaches (seemingly monotonically -- see the insets)  the value $\beta_+\leq 1/2$, thus providing evidence for the preservation of the linear version of SCC for angular momentum above the critical value $\acritq$.

\begin{figure}[!ht]
      \includegraphics[width=0.471\textwidth]{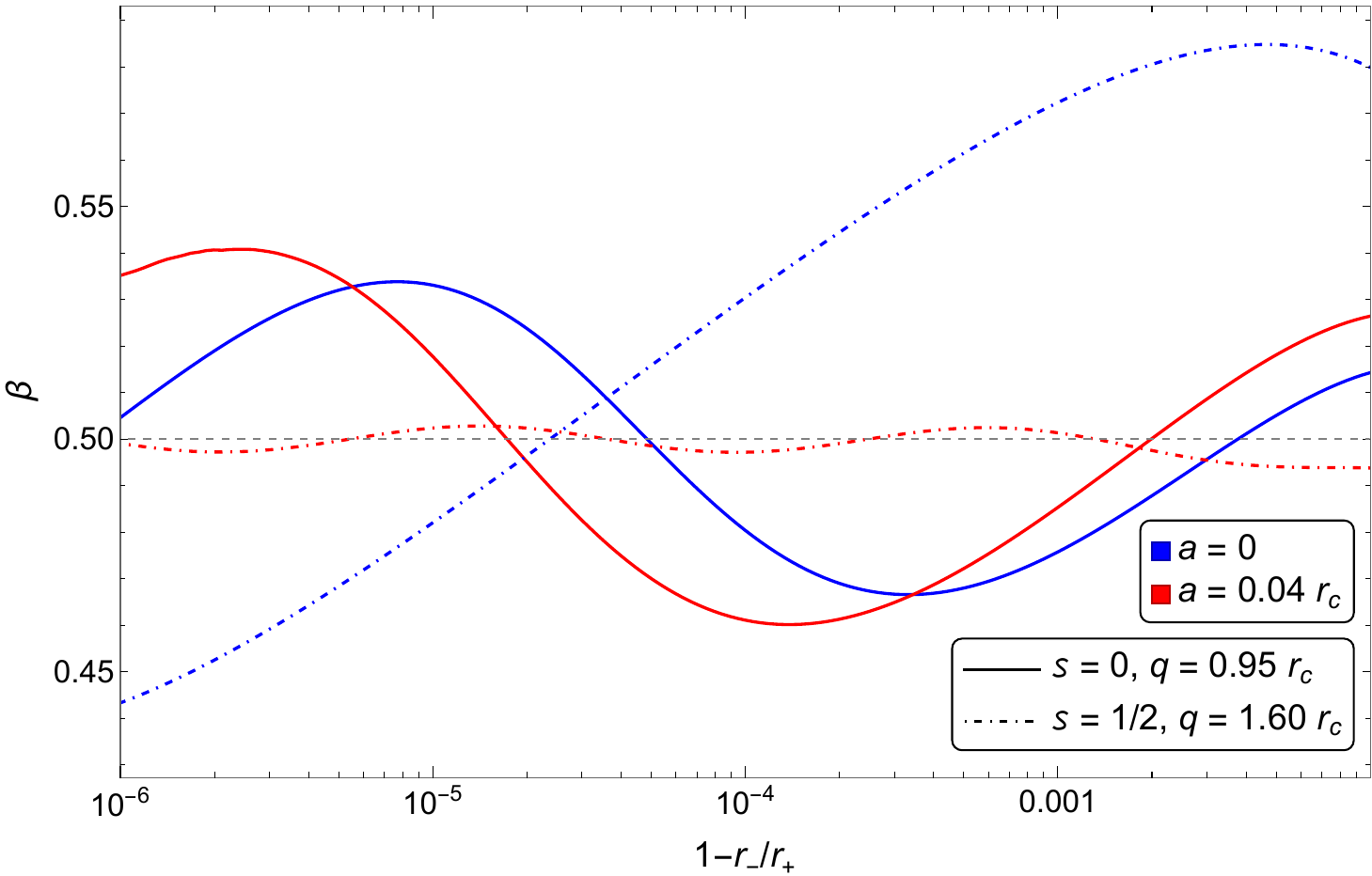}
    \caption{
    Plot of $\beta$ as a function of $1-r_-/r_+$ for $r_+=\rc/2$ for the cases: (i) $s=0$ and $q=0.95\rc$ (\PRDvtwo{solid}), and (ii) $s=1/2$ and $q=1.6\rc$ (dotdashed), both for $a=0$ (blue) and $a=0.04\rc$ \PRDvtwo{(red)}. The critical value $\beta=1/2$ is indicated as a horizontal \PRDvtwo{dashed} line.}
\label{fig:wiggles func sigma} 
\end{figure}

Let us now turn to the NE modes.
As $q$ increases from zero, $\beta_\text{NE}$ decreases until (approximately) the square root in Eq.~\eqref{eq:NE} becomes purely imaginary \PRDvtwo{--} let us denote by $q=\qcrit$ this critical value.
Then, for $q>\qcrit$, $\beta_\text{NE}$ displays ``wiggles"  around  $\beta_+$.
These wiggles are a non-perturbative effect missed by the WKB analysis and their amplitude decreases rapidly with $q$.
As a consequence, there is no severe $\beta\PRDvtwo{\geq}1$ (for spin-0) or $\beta\PRDvtwo{\geq}3/2$ (for spin-1/2) violation of the linear version of SCC for large $q$,  as opposed to what we observed for $q=0$.
On the other hand, the presence of the wiggles means that, for $a\lesssim \acrit$ and a given arbitrarily large value of $q$ (or at least over the significantly large region of values of $q$  where $\beta_{\text{PS}}>1/2$), 
one can find interval(s) of values  of $Q$ close enough to $Q_{\text{max}}$ in which $\beta>1/2$, as we shall demonstrate in Fig.~\ref{fig:wiggles func sigma} \footnote{We have further checked that, as $Q$ approaches $Q_{\text{max}}$,  the amplitude of the wiggles in Fig.~\ref{fig:wiggles func q}, as can be gathered from Fig.\ref{fig:wiggles func sigma}, does not decrease at a rate faster than the rate at which $1/2-\beta_+$ decreases.}.
As is clear from Fig.~\ref{fig:wiggles func q}, however, as $q$ increases, one would need more and more NE $(\ell,m)$-modes to synchronize their wiggles so that $\beta>1/2$ can be achieved.

The notable difference between the behaviour of the NE modes between spin-$0$ and spin-$1/2$ is that,
whereas $\beta_\text{NE}$ for $s=0$ increases with rotation (for fixed $q$) before the wiggles, it instead decreases for $s=1/2$.
Accordingly, whereas the wiggles start at a larger value of $q$ as rotation increases for $s=0$, they start at a smaller value of $q$  for $s=1/2$.
Thus, increasing rotation within $a\lesssim \acrit$ helps  achieve linear SCC violation $\beta>1/2$ for $s=0$ whereas it hinders violation for $s=1/2$.

\vspace{0.5cm}
{\it As a function of the black hole charge}
\vspace{0.5cm}

Last but not least, Fig.~\ref{fig:wiggles func sigma}  shows that wiggles also appear in $\beta$ as a function of $1-r_-/r_+$.
As may be inferred from Fig.~\ref{fig:wiggles func q}, the amplitude of the wiggles here increases with rotation for $s=0$ and decreases for $s=1/2$.
Since $\beta_+\to 1/2$ in the extremal limit, for $a\lesssim \acrit$ and a given field charge $q>\qcrit$, one can in principle find black holes sufficiently close to extremality  which have a $\beta>1/2$, thus violating the linear version of SCC.
This plot provides examples of regions of black hole parameter space where violation occurs when including both $s=0$ and $s=1/2$ charged fields.
Every time the wiggles of $\beta$ go above the value $1/2$, we catch a glimpse of SCC violation which, as we discussed for Fig.~\ref{fig:wiggles func q},  fades away as the field charge $q$ increases.

%------------------------------------------------------------------------------------------------
%------------------------------------------------------------------------------------------------

\section{Discussion}\label{sec:Discussion}
We have calculated the various families of QNMs for neutral and charged, massless scalar and fermion field perturbations of KNdS black hole spacetimes.
We have shown that there exist regions of phase space
where $\beta>1/2$ for both scalar and fermion fields, signaling the existence of weak solutions across the Cauchy horizon
(and, if only considering neutral fields, even $\beta>1$
for spin-0 and $\beta>3/2$ for spin-1/2, signaling boundedness of the curvature at the Cauchy horizon).
To the best of our knowledge, this is the first time that evidence for violation of SCC has been provided for a ($4$-dimensional) {\it rotating} black hole.

Regarding the parameter space considered, we have included black hole rotation and cosmological constant, which are important for modelling  black holes in Nature and the accelerated expansion of the Universe~\cite{Perlmutter:1998np,1998AJ....116.1009R}.
However, the large values of black hole charge required for SCC violation are astrophysically unrealistic~\cite{10.1093/mnras/179.3.433}.
As for the matter fields, considering them to be charged is required for the formation of a charged black hole. Again, physically realistic values correspond to $q\, r_c\gg 1$, where the glimpses of SCC violation fade away.
However, at least from a fundamental perspective, the evidence for SCC violation that we have shown remains disturbing\PRDvtwo{, since it means that  General Relativity would cease to be a predictive theory}.
From that perspective, two different options \PRDvtwo{for saving SCC that have been} proposed recently are probably worth investigating further: the inclusion of non-smooth initial data~\cite{Dafermos:2018tha} or of quantum effects (e.g.,~\cite{hollands2020quantum} in RNdS and~\cite{zilberman2022quantum} in Kerr).

%---------------------------------------------------------------------------------------------------------
%---------------------------------------------------------------------------------------------------------

\section{Acknowledgements}

Both M.C.~and C.M.~acknowledge partial financial support by CNPq (Brazil), Grants No.~310200/2017-2 and No.~314824/2020-0 in the case of M.C.~and Grant No.~142300/2017-9 in the case of C.M.

%---------------------------------------------------------------------------------------------------------
%---------------------------------------------------------------------------------------------------------

%---------------------------------------------------------------------------------------------------------
\appendix

\section{\texorpdfstring{QNM\lowercase{s}}{QNMS} AND SCC IN KERR-\texorpdfstring{\lowercase{d}S}{DS}}\label{sec:KdS}

\PRDvtwo{W}e also ran our QNM code in Kerr-dS (i.e., \PRDvtwo{KNdS with} $Q=0$) with two objectives. First, as a further check of our code\PRDvtwo{,} we checked that, for $s=-2$, we found agreement  to all digits with the QNMs in tables in Ref.~\cite{yoshida2010quasinormal} as well as visual agreement with their plots.

\begin{figure}[!h]
   \begin{center}
            \label{figZextreme2}
\includegraphics[width=.483\textwidth]{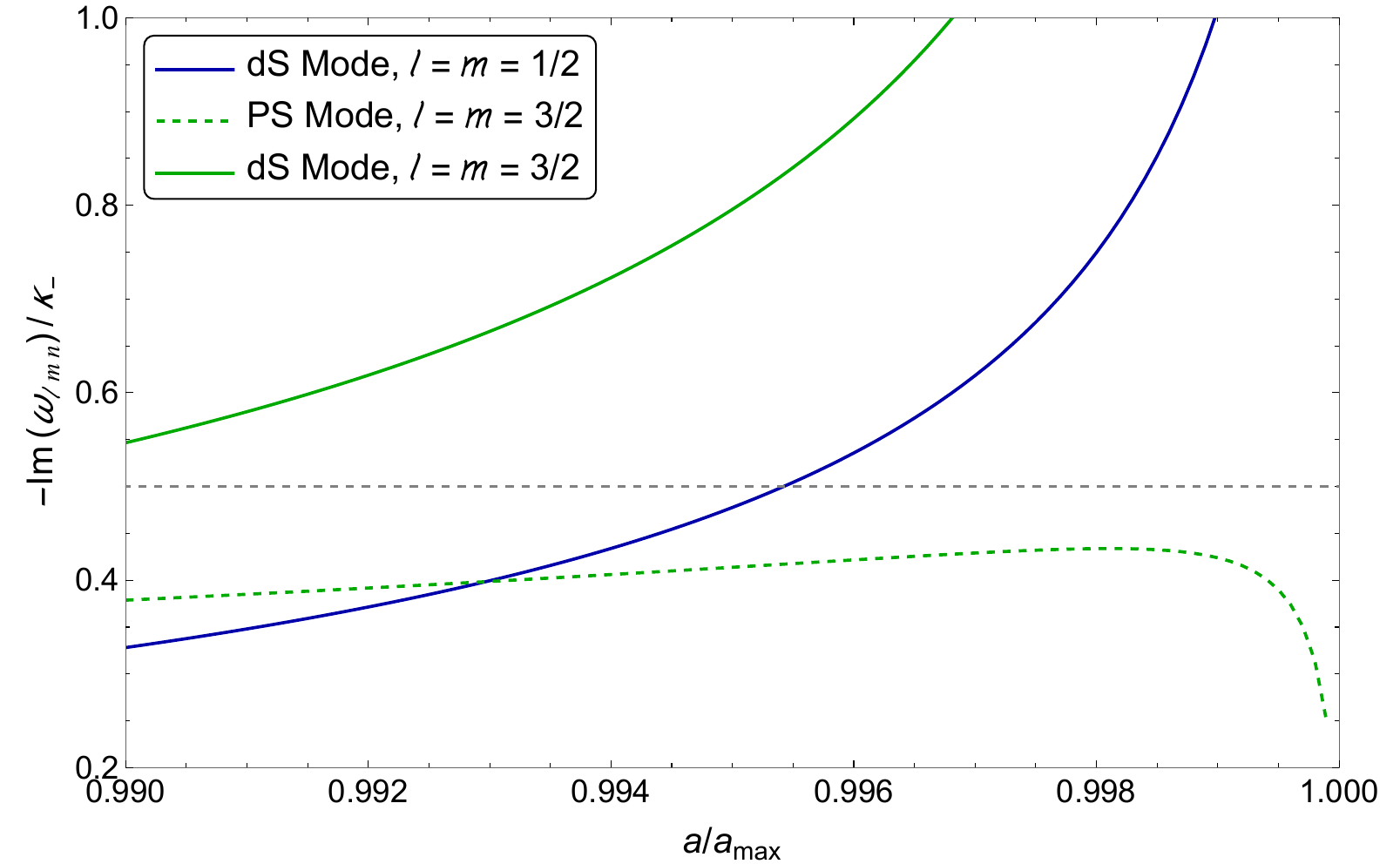}        
   \end{center}
\caption{
Plot of $-\text{Im}\left(\oQNM\right)/\kappa_-$ for the dominant modes of a (neutral) spin-1/2  field in Kerr-dS ($Q=0$) as a function of $a/a_{\text{max}}$ for $\Lambda M^2=0.001$.
The mode family and $(\ell,m)$ values are indicated in the inset.
Cf. Fig.2 (top left) in~\cite{mostafizur2020validity}.
}
\label{fig:KerrdS}
\end{figure}

The second objective is to investigate the claim in~\cite{mostafizur2020validity} that, in Kerr-dS, $\beta>1/2$ is possible for $s=1/2$.
In Fig.~\ref{fig:KerrdS} we plot the imaginary part of the most dominant QNM frequencies for spin-1/2 as a function of $a/a_{\text{max}}$ in the case of Fig.2 (top left) in~\cite{mostafizur2020validity},  where~\cite{mostafizur2020validity} claims violation of the linear version of
SCC.
Our curves for the dS modes for  $\ell=1/2$ and  $\ell=3/2$ agree with the curves 
in~\cite{mostafizur2020validity}. However, our Fig.~\ref{fig:KerrdS} shows that, while the slowest decaying mode for $\ell=1/2$ is indeed a dS mode, the slowest decaying mode for $\ell=3/2$  is instead a PS mode.
It seems that this mode is missed by~\cite{mostafizur2020validity} and, since it has $-\text{Im}\left(\oQNM\right)/\kappa_-<1/2$, it is in fact crucial for saving the linear version of SCC where~\cite{mostafizur2020validity} claims violation.
We have carried out similar comparisons for other values of black hole parameters (including the region in Fig.2 (top right) in~\cite{mostafizur2020validity} where  violation is claimed) and find the same outcome: Ref.~\cite{mostafizur2020validity} seems to follow the dS family only and other modes save  the linear version of SCC where~\cite{mostafizur2020validity} claims violation.

%---------------------------------------------------------------------
%---------------------------------------------------------------------

\section{EQUATORIAL CIRCULAR PHOTON ORBITS}\label{sec:geo_approx}

The large-$\ell$ behaviour of the QNMs can be described by the so-called eikonal limit or geometric optics approximation \cite{Yang:2012he}. In this approximation, there is a parallel between field waves and null geodesics, in  such a way that the real and imaginary parts of some QNM frequencies are associated with, respectively, the energy and the Lyapunov exponent of  unstable circular orbits of photons. The  Lyapunov exponent describes the exponential rate at which a perturbation of the radius of a circular photon orbit  grows or decays with time. 
In this appendix, we derive explicit analytic expressions for the energy and the Lyapunov exponent of the equatorial circular photon orbits in KNdS. These quantities yield the QNM frequencies for the PS family in Eq.~\eqref{eq:eikonal}.

The two Killing vectors $\boldsymbol{K}\equiv\partial_t$ and $\boldsymbol{M}\equiv \partial_\varphi$ yield two conserved quantities for a test particle on geodesics: $E~\equiv~-K_\mu\dot{x}^\mu$, $L_z~\equiv~M_\mu\dot{x}^\mu,$ where an overdot denotes differentiation with respect to an affine parameter $\zeta$. It follows from the symmetries of the spacetime that if the particle initially lies on the equatorial plane and has zero initial velocity along the $\theta$-direction, then the particle will remain at all times on that plane. That is, if $\theta(\zeta=\zeta_0) = \pi/2$ and
$\dot{\theta}(\zeta=\zeta_0)=0$ for some $\zeta_0\in\mathbb{R}$, then $\theta(\zeta) = \pi/2$, $\forall \zeta\in\mathbb{R}$. Henceforth we shall restrict ourselves to motion on the equatorial plane.

Let us now obtain the radial component of the tangent vector along null geodesics on the equatorial plane. This can readily be achieved from the null tangent vector condition (also using the definition of $E$ and $L_z$):
\begin{equation}
g_{\mu\nu}\dot{x}^\mu\dot{x}^\nu=g_{rr}\dot{r}^2-E\dot{t}+L_z\dot{\varphi}=0.\label{eq:r-geod2}
\end{equation}
We can rewrite $\dot{t}$ and $\dot{\varphi}$ in terms of $E,L_z,r$ and spacetime parameters $\{a,Q,M,L\}$ using our definitions for $E$ and $L_z$. By doing that, Eq.~\eqref{eq:r-geod2} can be written as
\begin{equation}\label{eq:r-geod3}
\dot{r}^2=
\frac{\Xi^2}{r^4}\left\{\left[E(r^2+a^2)-aL_z\right]^2-(L_z-aE)^2\Delta_r\right\}.
\end{equation}
\newcommand{\rph}{r_\text{ph}}
\newcommand{\Dph}{\Delta_{\text{ph}}}
\newcommand{\dDph}{\Delta'_{\text{ph}}}
\newcommand{\ddDph}{\Delta''_{\text{ph}}}

The radii $\rph^{\pm}$ of the  circular orbits are obtained from the conditions $\dot{r}=0$ and $\ddot{r}=0$. These two conditions together give us:
\begin{equation}\label{eq:energy_lz}
E=E^{\pm}\equiv
\frac{a\pm \sqrt{\Dph^\pm}}{\big(\rph^{\pm}\big)^2+a^2\pm a\sqrt{\Dph^\pm}}L_z,
\end{equation}
and
\begin{equation}\label{eq:rph}
    16a^2\Dr=\left(r\Dr'-4\Dr\right)^2,
\end{equation}
where $\Dph^\pm\equiv \Dr(\rph^\pm)$ and $\rph^{\pm}$ are the two largest (and real) roots of Eq.~\eqref{eq:rph}, with $\rph^-\geq\rph^+$. 
In this manner, $\rph^+$ is the radius of the co-rotating orbit and $\rph^-$ is the radius of the counter-rotating one.

Let us generically denote $\rph^\pm$ by $\rph$ and define $\Dph\equiv \Dr(\rph)$, $\dDph\equiv \Dr'(\rph)$ and $\ddDph\equiv \Dr''(\rph)$.
We now look for small deviations from the radii $\rph$. By setting $r(t)=\rph+\delta r(t)$, and expanding Eq.~\eqref{eq:r-geod3} in terms of $\delta r$, we have that, up to order $\mathcal{O}(\delta r(t)^3)$,
\begin{multline}\label{eq:dr'}
\dot{t}(\rph)^2\left[\delta r'(t)\right]^2=\frac{\Xi^2}{2\rph^4}\left[12\rph^2E^2-4aE(L_z-aE)\right.\\\left.-(L_z-aE)^2\ddDph\right]\delta r(t)^2+\mathcal{O}(\delta r(t)^3).
\end{multline}
We remind the reader that  a prime on a function means derivative with respect to its argument, so that, in particular, the prime in $\delta r'(t)$ means derivative with respect to $t$ and, in $\Dr'(r)$, with respect to $r$.
Ignoring the $\mathcal{O}(\delta r(t)^3)$, the general solution of Eq.~\eqref{eq:dr'} is
\begin{equation}
    \delta r(t) = c_1\exp(+ \LyapLambda t)+c_2\exp(- \LyapLambda t),
\end{equation}
where $c_{1,2}$ are constants and 
\begin{equation}\label{eq:lyapunov_ph}
\LyapLambda\equiv \frac{2}{\Xi}\sqrt{\frac{\rph(\dDph)^2+2\Dph(\dDph-\rph \ddDph)}{\rph\left[4(a^2+\rph^2)+\rph \dDph-4\Dph\right]^2}}
\end{equation}
is the so-called  Lyapunov exponent.

We now consider the case of near black hole extremality: $r_-=r_+-\epsilon\, L$, for small $\epsilon>0$. First, for small (but nonzero) $\epsilon$, Eq.~\eqref{eq:rph} has either two or four real roots depending on how the value $a_\text{max}$ of the angular momentum at the extremal black hole limit is related with the following critical value:
\begin{equation}
 \bar{a}_c \equiv r_+\sqrt{\frac{1-2\Lambda r_+^2}{4+\Lambda r_+^2/3}}.
\end{equation}
The value $\bar{a}_c/r_+$ is precisely equal to the value $\bar{a}_c^{\textrm{KNdS}}$ in Ref.~\cite{HOD2018221}. If $\bar{a}_c<a\lesssim a_\text{max}$, then there are four real roots, being  $\rph^+=r_+(=r_-)$ at the extremal black hole limit;  if $a\lesssim a_\text{max}<\bar{a}_c$, then  there are only two real roots and $\rph^+\neq r_+(=r_-)$ at the extremal black hole limit.

Now consider the case $\rph^+\to r_+$ as $\epsilon\to 0$ (i.e., $\bar{a}_c<a\lesssim a_\text{max}$).
Then, it can be shown that Eq.~\eqref{eq:lyapunov_ph} yields
\begin{equation}
    \LyapLambda=\frac{(r_c-r_+)(r_c+3r_+)\epsilon}{2L\Xi(r_+^2+a^2)} +\mathcal{O}(\epsilon^2)= \kappa_+ +\mathcal{O}(\epsilon^2).
\end{equation}
This means that, in this case, $\LyapLambda$ tends to $\kappa_+$, and therefore to zero, in the extremal black hole limit. Otherwise (i.e., for $a\lesssim a_\text{max}<\bar{a}_c$), this will not occur and $\LyapLambda$ tends to a non-zero value at the extremal black hole limit. Additionally, we have numerically checked that the quantity $\acrit$ defined in Sec.~\ref{sec:dominant} is such that $\acrit\to\bar{a}_c$ in the limit $\epsilon\to0$.

Finally, the eikonal approximation for the PS QNMs is obtained via the correspondence with geometric optics \cite{2009PhRvD..79f4016C,Yang:2012he}. 
So we map quantities of the circular photon orbits  to quantities of the PS QNM frequencies $\oPSn{}^\pm$  in the following manner: 
\begin{enumerate}
\item[(i)]  The energies
$E^{\pm}$ in \eqref{eq:energy_lz} are mapped to the real part of  $\oPSn{}^\pm$.
\item[(ii)]  Negative half-integer multiples of the Lyapunov exponent $\LyapLambda$ in \eqref{eq:lyapunov_ph} are mapped to the imaginary part of  $\oPSn{}^\pm$.
\item[(iii)] The azimuthal angular momentum $L_z$ appearing  in \eqref{eq:energy_lz} is mapped to
$m$ when $L_z$ is multiplied by $a$ and to $\text{sign}(m)(\ell+1/2)$ when it is not, where $\text{sign}(m)$ is 1 for $m\geq0$  and $-1$ for $m<0$. 
\end{enumerate}
Mappings (i) and (ii) are justified in \cite{2009PhRvD..79f4016C,Yang:2012he}.
In its turn,  mapping (iii), although more \textit{ad hoc}, is inspired by the fact that $L_z$ is mapped to $\pm(\ell+1/2)$ when $a=0$ \cite{2009PhRvD..79f4016C,Glampedakis:2019dqh} and to $m$ when $Q=0$ \cite{Yang:2012he,PhysRevD.97.104060}.
The result of such mappings is:
\begin{equation}\label{eq:opsn}
\oPSn{}^\pm =E_\ph^{\pm} -i\left(n+\frac{1}{2}\right)\LyapLambda^{\pm},\ \  \text{for} \quad n = 0,1,2,\dots
\end{equation}
where 
\begin{equation}\label{eq:Eph}
E_\ph^{\pm}\equiv
\frac{am+\text{sign}(m)(\ell+1/2)\sqrt{\Dph^\pm}}{\big(\rph^\pm\big)^2+a^2\pm a\sqrt{\Dph^\pm}},
\end{equation}
and $\LyapLambda^{\pm}$ is defined as the $\LyapLambda$ associated with $\rph^\pm$.
Offering further support for Eq.~\eqref{eq:opsn} is the fact that it reduces to the results of \cite{Tattersall:2018axd} for small rotation and to the results of \cite{PhysRevD.97.104060}  for $\ell=|m|\gg1$, both in Kerr-dS.
We emphasize that the imaginary part of $\oPSn{}^\pm$ in \eqref{eq:opsn}, which is the only relevant part for $\beta_{\text{PS}}$, does not depend on the mapping (iii) for $L_z$, and so it is to be regarded as the true value of the imaginary part of the QNM frequencies in the $\ell\to\infty$ limit. 
Only the real part of $\oPSn{}^\pm$ depends on this mapping for $L_z$, and we find it useful as seeds for numerically calculating the PS QNMs.

In order to reduce index cluttering in the main text, we define $\LyapLambda\equiv\LyapLambda^+$,
$E_\ph\equiv E_\ph^+$, and $\oPSn{}\equiv\oPSn{}^+$
as the  dominant PS mode is associated -- for the spacetime parameter values that we checked -- with the co-rotating equatorial photon orbit.

%---------------------------------------------------------------------------------------------------------
%---------------------------------------------------------------------------------------------------------

\section{THE MASSLESS CHARGED DIRAC EQUATION}\label{sec:app:Dirac}
Here we provide the necessary quantities for the analysis in Sec.~\ref{sec:beta} for charged fermion fields. In this appendix we  make use of the Newman-Penrose formalism  in
 the null tetrad of \cite{suzuki1998perturbations} after applying the coordinate transformation in our Eq.~\eqref{eq:outgoing-transf} to $\{u,r,\theta,\phio\}$ coordinates:
\begin{equation}\label{eq:tetrad_NP}
\begin{array}{rl}
\boldsymbol{l}=&\partial_r,\\[5pt]
\boldsymbol{n}=&\left(2\rho^2\right)^{-1}[2\Xi(r^2+a^2)\partial_u-\Dr\partial_r+2a\Xi\partial_{\phio}],\\[7pt]
\boldsymbol{m}=&\dfrac{1}{\bar{\rho}\sqrt{2\Delta_\theta}}\left[ia\Xi\sin\theta\partial_u+\Delta_\theta\partial_\theta+\dfrac{i\Xi}{\sin\theta}\partial_{\phio}\right]\!\!,\\[9pt]
\boldsymbol{\bar{m}} =&\dfrac{-1}{\bar{\rho}^*\sqrt{2\Delta_\theta}} \!\left[ia\Xi\sin\theta\partial_u-\Delta_\theta\partial_\theta+\dfrac{i\Xi}{\sin\theta}\partial_{\phio}\right]\!\!,
\end{array}
\end{equation}
where, $\bar\rho\equiv r+ia\cos\theta$.
We note that the null tetrad $\{\boldsymbol{l},\boldsymbol{n},\boldsymbol{m},\boldsymbol{\bar{m}}\}$ is regular on all horizons.
The metric can be written in terms of this null tetrad as $g^{\mu\nu} = -2l^{(\mu}n^{\nu)}+2m^{(\mu}\bar{m}^{\nu)}$. Now, the Dirac matrices $\gamma^\mu$, defined from $g^{\mu\nu}\mathbb{1}_4=\gamma^{(\mu}\gamma^{\nu)}$, can be expressed as
\begin{equation}\label{eq:gmunu_NP}
\gamma^\mu=i\sqrt{2}\left(\begin{matrix}0 & 0 & n^\mu & -\bar{m}^\mu\\ 0 & 0 & -m^\mu & l^\mu\\ l^\mu & \bar{m}^\mu & 0 & 0\\ m^\mu & n^\mu & 0 & 0\end{matrix}\right).
\end{equation}
Choosing the vierbein $e^\mu_I$, $I = \{1,2,3,4\}$, where $\boldsymbol{e}_1\equiv \boldsymbol{l}=-\boldsymbol{e}^2,\boldsymbol{e}_2\equiv \boldsymbol{n}=-\boldsymbol{e}^1,\boldsymbol{e}_3\equiv \boldsymbol{m}=\boldsymbol{e}^4,\boldsymbol{e}_4\equiv \boldsymbol{\bar{m}}=\boldsymbol{e}^3$, the spinor connection matrices $\Gamma_\mu$ are given by
\begin{equation}
\Gamma_\mu=-\frac{1}{4}e^K_\mu\gamma_{IJK}\gamma^I\gamma^J=-\frac{1}{4}e_{I\nu}e_{J;\mu}^\nu\gamma^I\gamma^J,
\end{equation}
where we have defined additional matrices $\gamma^I\equiv e_{\mu}^I\gamma^\mu$ and $\gamma_{IJK}$ are the Ricci rotation-coefficients \cite{Chandrasekhar}. Since our vierbein components in $\{u,r,\theta,\phio\}$ coordinates are all regular at $r=r_j$, $j=\{-,+,c\}$, and $\gamma_{IJK}$ are also  regular there, so are the spinor connection matrices $\Gamma_\mu$ in the same coordinates.

Let us now assume that the Dirac spinor is given by $\Psi_{1/2}=(F_1\ F_2\ -G_1\ -G_2)^T$, with $F_{1,2}$ and $G_{1,2}$ functions of $\{u,r,\theta,\phio\}$. After performing the gauge transformation $\boldsymbol{A}\to \boldsymbol{A}+d\chi$ and $\Psi_{1/2}\to e^{iq\chi}\Psi_{1/2}$ with $d\chi=Qrdr/\Dr$,  the Dirac equation \eqref{eq:DiracEquation} can be written out explicitly as
\begin{align}\label{eq:eqs F,G}
0=&\left(l^\mu\partial_\mu+\frac{1}{\bar{\rho}^*}\right)F_1+\left(\bar{m}^\mu\partial_\mu+\beta_\text{NP}^*\right)F_2,\\
0=&\left[n^\mu\partial_\mu-\frac{\Dr'}{4\rho^2}+\frac{iqQr}{\rho^2}\right]\!F_2+\left(m^\mu\partial_\mu+\beta_\text{NP}-\tau_\text{NP}\right)\!F_1,\nonumber\\
0=&\left(l^\mu\partial_\mu+\frac{1}{\bar{\rho}}\right)G_2-\left(m^\mu\partial_\mu+\beta_\text{NP}\right)G_1,\nonumber\\
0=&\left[n^\mu\partial_\mu-\frac{\Dr'}{4\rho^2}+\frac{iqQr}{\rho^2}\right]\!G_1-\left(\bar{m}^\mu\partial_\mu+\beta_\text{NP}^*-\tau_\text{NP}^*\right)\!G_2,\nonumber
\end{align}
where
\begin{equation}
\beta_\text{NP}\equiv\frac{\partial_\theta(\sin\theta\sqrt{\Delta_\theta})}{2\sqrt{2}\bar{\rho}\sin\theta}, \quad \tau_\text{NP} \equiv \frac{-ia\sin\theta \sqrt{\Delta_\theta}}{\sqrt{2}\rho^2}.
\end{equation}
The metric is  independent of the coordinates $u$ and $\phio$, corresponding to the existence of the Killing vectors $\partial_u$ and $\partial_{\phio}$. This allows us to separate $F_{1,2}$ and $G_{1,2}$ into  $u$- and $\phio$-dependent mode factors  as $e^{i(m\phio-\omega u)}$. We can further separate these spinor components  into $r$- and $\theta$-dependent factors yielding the following mode decomposition: 
%\begin{widetext}
%\begin{equation}\label{eq:Psihalfspin}
%\Psi_{1/2}=\left(\!\begin{array}{c}
%    F_1\\ F_2\\ - G_1\\ -G_2 \end{array}\!\right)
%    =\int_{\mathbb{R}}d\omega\sum_{\ell=1/2}^{\infty}\sum_{m=-\ell}^{\ell}
%{}_{1/2}\tilde{c}_{\indmode}e^{i(m\phio-\omega u)}
%\left(\begin{array}{c}
%    \tilde{R}_{-1/2}(r)S_{-1/2}(\theta)/(\bar{\rho}^*\sqrt{2})\\
%    \tilde{R}_{+1/2}(r)S_{+1/2}(\theta)\\
%    -\tilde{R}_{+1/2}(r)S_{-1/2}(\theta)\\
%    -\tilde{R}_{-1/2}(r)S_{+1/2}(\theta)/(\bar{\rho}\sqrt{2})
%\end{array}\right),
%\end{equation}
%\end{widetext}

\begin{align}\label{eq:Psihalfspin}
\Psi_{1/2}&=\left(\!\begin{array}{c}
    F_1\\ F_2\\ - G_1\\ -G_2 \end{array}\!\right)
    =&\\
    &=\int_{\mathbb{R}}d\omega\sum_{\ell=1/2}^{\infty}\sum_{m=-\ell}^{\ell}
{}_{1/2}\tilde{c}_{\indmode}e^{i(m\phio-\omega u)}\PRDvtwo{\tilde\psi_{\indmode}(r,\theta),}\nonumber
\end{align}
where
\PRDvtwo{\begin{equation}
\tilde\psi_{\indmode}(r,\theta)\equiv    \left(\begin{array}{c}
    \tilde{R}_{-1/2}(r)S_{-1/2}(\theta)/(\bar{\rho}^*\sqrt{2})\\
    \tilde{R}_{+1/2}(r)S_{+1/2}(\theta)\\
    -\tilde{R}_{+1/2}(r)S_{-1/2}(\theta)\\
    -\tilde{R}_{-1/2}(r)S_{+1/2}(\theta)/(\bar{\rho}\sqrt{2})
\end{array}\right),
\end{equation}}\\
\noindent\PRDvtwo{and} ${}_{1/2}\tilde{c}_{\indmode}$ are complex constants. With $\Psi_{1/2}$ given by Eq.~\eqref{eq:Psihalfspin} satisfying \eqref{eq:eqs F,G}, one can directly check that $S_{\pm1/2}$ satisfy Eq.~\eqref{eq:angulareq} for $s=\pm1/2$ and $R_{s} = e^{-iq\chi} e^{i\omega r_*} e^{-im r_\phi} \tilde{R}_{s}$, with $dr_\phi\equiv \frac{a \Xi}{\Delta_r}dr$ [see Eq.~\eqref{eq:outgoing-transf}], satisfy Eq.~\eqref{eq:radialeq} for $s=\pm1/2$. 

We finish this appendix by writing the equation for the conjugate spinor $\bar\Psi_{1/2}\equiv\Psi_{1/2}^\dagger\bar\gamma^0$, where $\dagger$ denotes the usual Hermitian conjugation. A suitable choice of a matrix $\bar\gamma^0$ that has the properties
\begin{equation}
\gamma^{\mu\dagger}\bar\gamma^0+\bar\gamma^0\gamma^{\mu}=0 \quad \textrm{and} \quad \partial_\mu \bar\gamma^0+ \Gamma_\mu^\dagger\bar\gamma^0+\bar\gamma^0\Gamma_\mu=0,
\end{equation}
is
\begin{equation}\label{eq:bargamma0}
\bar\gamma^0=-i\left(\begin{matrix}\mathbb{0} & \mathbb{1}_2 \\ \mathbb{1}_2 & \mathbb{0} \end{matrix}\right).
\end{equation}
Then the conjugate spinor $\bar\Psi_{1/2}$ will satisfy the equation
\begin{equation}
\left(D_\mu\bar\Psi_{1/2}\right)\gamma^\mu = (\partial_\mu\bar\Psi_{1/2}+\Gamma_\mu\bar\Psi_{1/2}+iqA_\mu\bar\Psi_{1/2})\gamma^\mu=0.
\end{equation}

\end{document}